\documentclass[%
 aip,
 amsmath,amssymb,
preprint,%
]{revtex4-1}

\usepackage{graphicx}
\usepackage{dcolumn}
\usepackage{bm}

\usepackage[utf8]{inputenc}
\usepackage[T1]{fontenc}
\usepackage{mathptmx}
\usepackage{etoolbox}
\usepackage{graphicx}
\graphicspath{ {./images/} }
\usepackage{epstopdf, epsfig}
\usepackage{amsmath,amssymb}
\usepackage{float}
\usepackage{subfig}
\usepackage{upgreek}
\usepackage{bm}
\usepackage{mathtools}  
\usepackage{diffcoeff}  
\usepackage[dvipsnames]{xcolor}
\usepackage{hyperref}
\usepackage{comment}

\makeatletter
\def\@email#1#2{%
 \endgroup
 \patchcmd{\titleblock@produce}
  {\frontmatter@RRAPformat}
  {\frontmatter@RRAPformat{\produce@RRAP{*#1\href{mailto:#2}{#2}}}\frontmatter@RRAPformat}
  {}{}
}%
\makeatother

\begin{document}


\title[Droplet Migration with a Reacting Solute]{Droplet Migration in the Presence of a Reacting Surfactant at Low P\'eclet Numbers}

\author{S. Roychowdhury}
\author{R. Chattopadhyay}%
 \affiliation{Department of Chemical Engineering, Jadavpur Univeristy, Kolkata, 700032, India}

\author{R. Mangal}
\author{D. S. Pillai}
\email{dipinsp@iitk.ac.in}
\affiliation{
Department of Chemical Engineering, Indian Institute of Technology Kanpur, Uttar Pradesh, 208016, India
}

\date{\today}

\begin{abstract}
A surfactant--laden droplet of one fluid dispersed in another immiscible fluid serves as an artificial model system capable of mimicking microbial swimmers. Either an interfacial chemical reaction or the process of solubilization generates gradients in interfacial tension resulting in a Marangoni flow. The resulting fluid flow propels the droplet toward a region of lower interfacial tension. The advective transport of surfactants sustains the active propulsion of these droplets. In these systems, the local interfacial tension is affected by the interfacial reaction kinetics as well as convection and diffusion induced concentration gradients. The migration of such a surfactant-laden viscous droplet undergoing an interfacial reaction, suspended in a background Poiseuille flow is investigated. The focus is specifically on the role of the surface reaction that generates a non-uniform interfacial coverage of the surfactant, which in turn dictates the migration velocity of the droplet in the background flow. Assuming negligible interface deformation and fluid inertia, the Lorentz reciprocal theorem is used to analytically determine the migration velocity of the droplet using regular perturbation expansion in terms of the surface P\'eclet number. We show that the presence of interfacial reaction affects the magnitude of both stream-wise and cross-stream migration velocity of the droplet in a background Poiseuille flow. We conclude that the stream-wise migration velocity is not of sufficient strength to exhibit positive rheotaxis as observed in recent experimental observations. Additional effects such as the hydrodynamic interactions with the adjacent wall may be essential to capture the same.

\end{abstract}

\maketitle

\section{Introduction}
Active droplets or swimming droplets \citep{maass2016swimming} are droplets that propel themselves into motion when dispersed in a surfactant-laden ambient medium.  Self-propelled motion of droplets may be generated either via an interfacial chemical reaction that alters the surfactant activity \citep{Thutupalli2011} or due to the process of micellar solubilization \citep{peddireddy2012solubilization}. In either case, the erstwhile spherical symmetry is spontaneously broken and a self-sustained propulsion state is achieved. These swimming droplets show promising potential in understanding the solitary as well as collective behaviour of simple micro-organisms \citep{yoshinaga2017simple}. They exhibit rich dynamics owing to the two-way nonlinear coupling between the surfactant chemical field and the fluid velocity fields, resulting in interesting phenomena such as autochemotaxis \citep{jin2017chemotaxis} and rheotaxis \cite{rheotaxis_of_active-droplets}.

Surfactants are molecules that preferentially adsorb at the liquid interface, thereby reducing the interfacial tension. A nonuniform distribution of surfactants can result in an unbalanced interfacial stresses which causes motion. This kind of flow induced by a gradient in surface tension at the interface of liquids is called the Marangoni effect \citep{young1959motion,levich1962motion}. This Marangoni effect is at the heart of the self-propulsion mechanism, and in the presence of a background flow, influences both the stream-wise flow and cross streamline migration. \citet{Hanna2010} investigated the migration of a non-deforming drop induced by a gradient in surfactant concentration, with an unbounded Poiseuille flow in the background, considering high values of Marangoni number and viscosity ratio. The drop was found to exhibit transverse migration towards the center of the Pouseuille flow due to change in surfactant distribution. Later \citet{pak2014viscous} investigated the migration of an insoluble surfactant laden spherical drop suspended in an unbounded Poiseuille flow in the limits of Stokes flow and low surface P\'eclet number. They showed that surfactant redistribution by the background flow retards the droplet motion, and this slip velocity was found to be independent of the droplet's position in the Poiseuille flow. They further noticed a cross stream migration of the drop towards the center of the Poiseuille flow, with the magnitude of migration velocity being linearly dependent on the distance of the drop from the center of the flow. \citet{mandal2015effect} analyzed the motion of a surfactant-free deforming droplet suspended in an unbounded Poiseuille flow in the limit of Stokes flow. They showed that the droplet can migrate towards or away from the centerline of the channel depending on the viscosity ratio. 

An active droplet may exhibit self-propulsion either due to solubilization or due to interfacial chemical reaction. We model such an active droplet as a surfactant-laden, non-deforming droplet undergoing a first-order interfacial reaction at the interface that alters the surfactant activity. Such a simplistic model was shown by Thutupalli et al. \cite{Thutupalli2011} to be able to capture the self-propelling dynamics of a droplet driven by interfacial chemical reaction. We theoretically determine the migration velocity for such a reacting droplet undergoing a first-order reaction at the interface in a background linear shear flow. The objective is to determine the role of interfacial reaction and the resultant surfactant redistribution on the droplet migration velocity, and to investigate the possibility of this mathematical model to predict the recent experimentally observed positive rheotactic phenomenon \cite{rheotaxis_of_active-droplets}. In these experiments, it has been observed that a gravitationally settled active droplet, in the presence of pressure-driven flow, propels in the direction opposite to the flow showing positive rheotaxis. It is theorised that the upstream rheotactic behaviour may be either due to the effect of the background linear flow or hydrodynamic interactions with the bottom wall or both.

We extend the work of \citet{pak2014viscous} to investigate the role of the interfacial reaction that alters surfactant activity on the resultant droplet migration velocity in an unbounded Poiseuille flow in the limits of low surface P\'eclet number, Such a reacting droplet is relevant in the case of swimming droplets propelled due to interfacial reaction such as an aqueous droplet propelling in monoolein-laden squalane due to bromination of monoolein  \citep{Thutupalli2011,tanabe2020effect}. We employ the Lorentz reciprocal theorem to  avoid tedious and involved calculations for the full solution to the velocity fields.

\section{Mathematical model}\label{sec:mathmodel}
\subsection{Description of the physical problem}\label{sec:phyprob}
\begin{figure}
    \centering
    \includegraphics[scale=0.5]{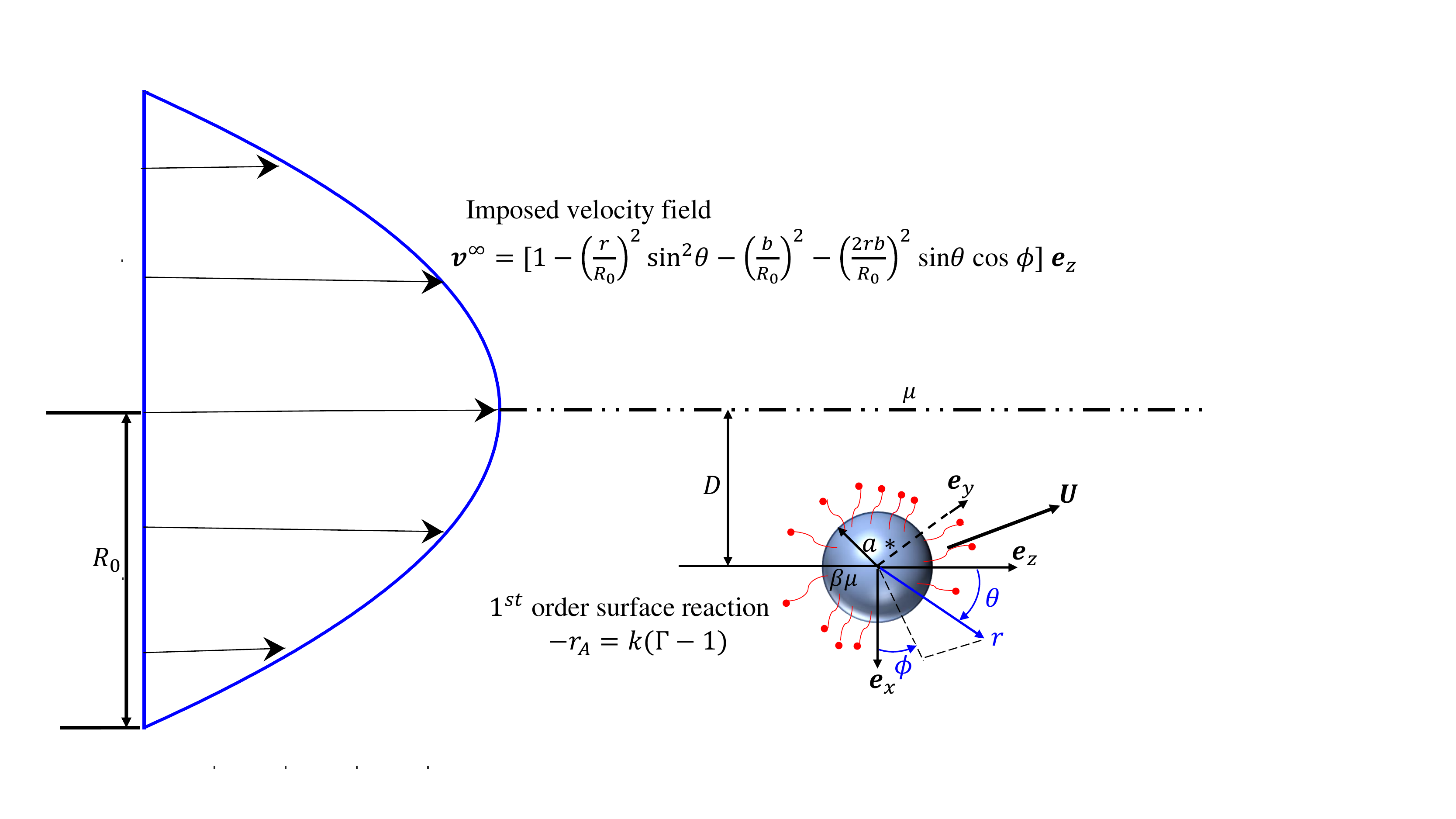}
    \caption{Sketch of the problem: a neutrally buoyant Newtonian drop with a reacting solute at the interface suspended in another Newtonian medium with a background Poiseuille flow, at distance $D$ from the flow centre line.
} \label{fig:Schematic}
\end{figure}

A neutrally buoyant Newtonian droplet of radius $a^*$ and viscosity $\beta\mu$, is placed in an ambient Newtonian fluid with viscosity, $\mu$ as considered in figure \ref{fig:Schematic}. Here, $\beta$ is the viscosity ratio of drop to that of the ambient medium. The suspending liquid undergoes an unbounded Poiseuille flow far away from the drop, the strength of which is characterized by the velocity scale, $U_b$. A rectangular coordinate system is defined with its origin at the centre of the drop, where $z$ is along the axis of the Poiseuille flow, $x$ is along the width of the flow but on the same plane as $z$, and $y$ is out of the plane. The Reynolds numbers for both the droplet and the ambient phase are assumed to be sufficiently small such that the viscous forces dominate, and inertia may be neglected completely. The capillary number, $\mu U_{b}/\upgamma_c$, is also assumed to be very low so that the droplet maintains its spherical shape at all times.The aqueous droplet interface is covered with an adsorbed layer of surfactant which is continuously undergoing a surface reaction following a first order kinetics affecting the interfacial tension at the droplet surface, augmenting the convection and diffusion induced concentration distribution. This results in a self-sustained surfactant concentration gradient along the droplet-suspending fluid interface, which propels the droplet due to Marangoni stresses \citep{Thutupalli2011}. The interfacial tension, $\upgamma^*$, is considered to have a linear dependence on the surfactant concentration, $\Gamma^*$, related as $\upgamma^{*} =\upgamma_{c}-RT\Gamma^*$, where $\upgamma_c$ is the interfacial tension for a clean interface. This relationship stands true for sufficiently dilute surfactant concentration \citep{Pawar1996, Adamson1997}. 

\subsection{Non-dimensionalization}\label{sec:nondim}
We first introduce the following scales to non-dimensionalize the variables denoted with an asterisk. 
\begin{gather*}
    x = \frac{x^*}{a^*},\quad y= \frac{y^*}{a^*},\quad z=\frac{z^*}{a^*}, \quad R_0=\frac{R_{0}^*}{a^*}, \quad a=\frac{a^*}{a^*} \quad \bm{v}=\frac{\bm{v}^*}{U_b}, \quad \hat{\bm{v}}=\frac{\hat{\bm{v}}^*}{U_b}, \quad \Gamma=\frac{\Gamma^*}{\Gamma_{eq}}, \\ \quad p=\frac{p^*}{\mu U_{b}/a^*},
    \quad \hat{p}=\frac{\hat{p}^*}{\beta\mu U_{b}/a^*}, \quad \bm{\sigma}=\frac{\bm{\sigma}^*}{\mu U_{b}/a^*}, \quad \hat{\bm{\sigma}}=\frac{\hat{\bm{\sigma}}^*}{\beta\mu U_{b}/a^*}
\end{gather*}
We have non-dimensionalized lengths by the drop radius $a^*$; velocity outside the drop, $\bm{v}$, and inside the drop, $\hat{\bm{v}}$, by the specified velocity  of the background Poiseuille flow, $U_b$; and surfactant concentration by its value at equilibrium when the distribution is uniform, $\Gamma_{eq}$. The pressure and stress fields outside the drop, $\left(p,\bm{\sigma}\right)$, and inside the drop $\left(\hat{p},\hat{\bm{\sigma}}\right)$, are non-dimensionalised by $\mu U_{b}/a^*$ and $\beta\mu U_{b}/a^*$ respectively. All the variables used subsequently are to be considered dimensionless unless otherwise mentioned.
\subsection{Velocity field and boundary conditions}\label{sec:velfield}
As $Re\ll1$, for both the droplet and suspending fluid, we can safely assume that the pressure and velocity fields inside the drop, $\left(\hat{p},\hat{\bm{v}}\right)$,and, outside the drop, $\left(p,\bm{v}\right)$, will satisfy the Stokes equation  and continuity equation
\begin{subequations}  
\begin{equation}
     \nabla p=\nabla^2 \bm{v}, \quad \nabla \cdot \bm{v}=0, 
\end{equation}
\begin{equation}
     \nabla \hat{p}=\nabla^2 \hat{\bm{v}}, \quad \nabla \cdot \hat{\bm{v}}=0.
\end{equation}
\end{subequations}
The velocity field far away from the droplet approaches the unperturbed background flow velocity, $\bm{v}^\infty$:
\begin{equation}
    \bm{v}\left(r\rightarrow\infty\right)=\bm{v}^{\infty}.
\end{equation}
The background flow velocity, $\bm{v}^\infty$, and the undisturbed pressure field, $p^\infty$, also satisfies the Stokes equations.\\
The kinematic boundary conditions at the interface $\left(r=1\right)$, is given by 
\begin{subequations}
\begin{align}
    \bm{v}&=\hat{\bm{v}},\\
    \bm{n}\cdot\bm{v}&=\bm{n}\cdot\bm{U},
\end{align}
\end{subequations}
where $\bm{U}$ is the migration velocity of the droplet and $\bm{n}=\bm{e}_r$ is the unit normal vector directed outwards on the spherical surface of the drop.\\
The non dimensional form for the tangential stress balance along the droplet interface is given by
\begin{equation}\label{2.4}
    \bm{n}\cdot\left(\bm{\sigma}-\beta\hat{\bm{\sigma}}\right)\cdot\left(\bm{I}-\bm{nn}\right)=Ma\nabla_{s}\Gamma,
\end{equation}
where $\nabla_{s}$, the surface gradient operator is given as, $\nabla_{s}=\left(\bm{I}-\bm{nn}\right)\cdot\nabla$ . The relationship that has been used to associate the interfacial tension and the surfactant concentration with the Marangoni number in $\left(\ref{2.4}\right)$, has been defined as
\begin{equation}
    Ma=\frac{RT\Gamma_{eq}}{\mu U_{b}},
\end{equation}
which compares the viscous stress to the concentration gradients of the surfactant.
\subsection{Surfactant transport equation}\label{sec:Tpeqns}
The surfactant transport is governed by the convection-diffusion equation coupled with a first--order reaction going on along the droplet surface. Further analytical progress was done considering a quasi-steady state \citep{Hanna2010}, which gives the following  transport equation
\begin{equation}\label{2.6}
    Pe_{s}[\nabla_{s}\cdot\left(\Gamma\bm{v}_{s}\right)+\Gamma\left(\bm{v}\cdot\bm{n}\right)\nabla_{s}\cdot\bm{n}]+k\left(\Gamma-1\right)=\nabla^2\Gamma,
\end{equation}
where $\bm{v}_s$ is the tangential component of the velocity along the surface of the drop and $k$ is the dimensionless 1\textsuperscript{st} order rate constant defined as, $k=k^*\times\frac{{a^*}^2}{D_{s}}$, where $k^*$ is the dimensional rate constant. Also,
\begin{equation}
    Pe_{s}=\frac{aU_{b}}{D_{s}}
\end{equation}
is the surface P\'eclet number which signifies the ratio of the rate of convection to the rate of diffusion, where $D_{s}$ is the  surface-diffusion coefficient of the surfactant. In equation $\left(\ref{2.6}\right)$, the first term on the LHS signifies the convective contributions to the surfactant transport, the second term represents the changes in the concentration of the surfactant induced by local variations in interfacial area \citep{Stone1990} and the third term shows concentration variation due to the reaction of the solute along the drop surface. The term on the right-hand side of $\left(\ref{2.6}\right)$, accounts for the transport of surfactants due to diffusion. It is important to mention, that rate constant $k$ is independent from the influence of  P\'eclet number.
\subsection{A moving reference frame approach}\label{sec:movref}
In our model, the reference frame is assumed to be moving with the unknown  migration velocity of the drop, $\bm{U}$, such that the drop appears to be stationary. In this moving frame, the velocity fields outside and inside the drop are given by $\left(\bm{u},\hat{\bm{u}}\right)=\left(\bm{v}-\bm{U},\hat{\bm{v}}-\bm{U}\right)$, which similarly satisfies the Stokes and continuity equations
\begin{subequations}\label{2.8}
\begin{equation}
     \nabla p=\nabla^2 \bm{u}, \quad \nabla \cdot \bm{u}=0 ,
\end{equation}
\begin{equation}
     \nabla \hat{p}=\nabla^2 \hat{\bm{u}}, \quad \nabla \cdot \hat{\bm{u}}=0.
\end{equation}
\end{subequations}
In the far-field $\left(r\rightarrow\infty\right)$, the velocity is now given by
\begin{equation}\label{2.9}
    \bm{u}\left(r\rightarrow\infty\right)=\bm{u}^{\infty}=\bm{v}^{\infty}-U.
\end{equation}
The kinematic boundary conditions at the interface $\left(r=1\right)$, are now given by
\begin{subequations}
\begin{align}
    \bm{u}&=\hat{\bm{u}},\\
    \bm{n}\cdot\bm{u}&=0.
\end{align}
\end{subequations}
The tangential stress balance remains unaltered and retains its form as in $\left(\ref{2.4}\right)$ as the stress fields and the surfactant concentration remains constant with the change in the reference frames.\\
Since $\bm{n}\cdot\bm{u}=0$ is true and the drop surface remains same in both the reference frames, the surfactant transport equation $\left(\ref{2.6}\right)$ becomes
\begin{equation}\label{2.11}
     Pe_{s}\nabla_{s}\cdot\left(\Gamma\bm{u}_{s}\right)+k\left(\Gamma-1\right)=\nabla^2\Gamma,
\end{equation}
where now $\bm{u_s}$ is the tangential velocity along the drop surface in the new reference frame.\\
The yet-to-be determined droplet migration velocity $\bm{U}$ appearing in $\left(\ref{2.9}\right)$ is the primary objective of this paper. In the next section, an analytical solution for $\bm{U}$ as a function of $Pe_{s}$ and $Ma$ has been derived using reciprocal theorem.
\section{A reciprocal theorem approach}\label{sec:recthrm}
\subsection{Translational  velocity of the droplet}\label{sec:genan}
To make our reciprocal theorem approach more convenient we represent the perturbed flows in both the drop fluid ($\hat{\bm{u}}'$) and the suspending fluid ($\bm{u}'$), relative to the background flow velocity far from the drop($\bm{u}^\infty$) as done by Shun Pak and Stone \citep{pak2014viscous}
\begin{subequations}

\begin{equation}
    \bm{u}'=\bm{u}-\bm{u}^\infty,
\end{equation}
\begin{equation}
    \hat{\bm{u}}'=\hat{\bm{{u}}}-\bm{u}^\infty.
\end{equation}
\end{subequations}
As $r$ approaches far field, the disturbance in the suspending fluid decays rapidly to become insignificant
\begin{equation}
    \bm{u}'(r\to\infty)=0.
\end{equation}
The new boundary conditions are:
\begin{subequations}
\begin{align}
    \bm{u}'&=\hat{\bm{u}}',\label{3.3a}\\
    \bm{n\cdot u}'&=-\bm{n}\cdot \bm{u}^\infty.\label{3.3b}
\end{align}
\end{subequations}
Considering the new frame of reference we define the stress fields of the flow inside and outside the drop as $\hat{\bm{\sigma}}'$ and $\bm{\sigma}'$ respectively. The stress field of the background flow is denoted by $\bm{\sigma}^\infty$. So the revised tangential stress balance equation $(\ref{2.4})$ becomes
\begin{equation}\label{3.4}
 \bm{n}\cdot (\bm{\sigma}'-\beta\hat{\bm{\sigma}}')\cdot (\bm{I}-\bm{nn})=(\beta -1)\bm{n}\cdot \bm{\sigma}^\infty\cdot (\bm{I}-\bm{nn})+Ma\nabla_s\Gamma. 
\end{equation}
We need to consider a supplemental problem to implement the reciprocal theorem. So we use the Hadamard-Rybczynski problem, which consists of a uniform background flow $(\bm{U}_{aux})$ of a viscous fluid past a clean and stationary drop that is spherical in shape, as our auxiliary problem, and define its stress fields and velocity outside the drop as $(\bm{\sigma}_{aux},\bm{u}_{aux})$ and inside the drop as $(\hat{\bm{\sigma}}_{aux},\hat{\bm{u}}_{aux})$ respectively. It is known that both the auxiliary problem and the main problem satisfy Stokes equation, so we can implement reciprocal theorem as done by \citep{leal1980particle, nadim1990thermocapillary,rallison1978note}.
The reciprocal theorem when applied outside the drop, we get
\begin{equation}\label{3.5}
    \int_\infty\bm{n}\cdot\bm{\sigma}'\cdot\bm{u}_{aux}\text{d}S-\int_S\bm{n}\cdot\bm{\sigma}'\cdot\bm{u}_{aux}\text{d}S = \int_\infty\bm{n}\cdot\bm{\sigma}_{aux}\cdot\bm{u}'\text{d}S-\int_S\bm{n}\cdot\bm{\sigma}_{aux}\cdot\bm{u}'\text{d}S.
\end{equation}
Applying reciprocal theorem inside the drop we get
\begin{equation}\label{3.6}
    \int_S\bm{n}\cdot\hat{\bm{\sigma}}'\cdot\hat{\bm{u}}_{aux}\text{d}S=\int_S\bm{n}\cdot\hat{\bm{\sigma}}_{aux}\cdot\hat{\bm{u}}'\text{d}S.
\end{equation}
Here $S$ denotes spherical drop surface, $\infty$ is an imaginary spherical surface far away from the drop and $\bm{n}=\bm{e}_r$ is the unit vector normal to the spherical surface. In the first term on the left hand side of $(\ref{3.5})$, we have $\bm{u}_{aux}=\bm{U}_{aux}$ at infinity and hence can be taken out of the integral making the whole term zero since the drop does not experience any force i.e., $ \int_\infty\bm{n}\cdot\bm{\sigma}'\cdot\bm{u}_{aux}dS= \bm{U}_{aux}\cdot\int_\infty\bm{n}\cdot\bm{\sigma}'dS=0$. The first term in the RHS of $(\ref{3.5})$ also becomes zero since the product of stress in the auxiliary problem and disturbance velocity decays at a rate higher than $1/r^2$ in the far field. So $(\ref{3.5})$ becomes
\begin{equation}\label{3.7}
    \int_S\bm{n}\cdot\bm{\sigma}'\cdot\bm{u}_{aux}\text{d}S = \int_S\bm{n}\cdot\bm{\sigma}_{aux}\cdot\bm{u}'\text{d}S.
\end{equation}
$(\ref{3.6})$ is multiplied by $\beta$ and the product is subtracted from $(\ref{3.7})$ to get
\begin{equation}\label{3.8}
     \int_S\bm{n}\cdot(\bm{\sigma}'-\beta\hat{\bm{\sigma}}')\cdot\bm{u}_{aux}\text{d}S = \int_S\bm{n}\cdot(\bm{\sigma}_{aux}-\beta\hat{\bm{\sigma}}_{aux})\cdot\bm{u}'\text{d}S.
\end{equation}
Applying tangential stress balance $(\ref{3.4})$ on the left hand side term and decomposing $\bm{u}'$ in the right hand side term to normal and tangential components in $(\ref{3.8})$ 
\begin{equation}\label{3.9}
    \int_S[(\beta -1)\bm{n}\cdot \bm{\sigma}^\infty\cdot (\bm{I}-\bm{nn})+Ma\nabla_s\Gamma]\cdot\bm{u}_{aux}\text{d}S=\int_S\bm{n}\cdot(\bm{\sigma}_{aux}-\beta\hat{\bm{\sigma}}_{aux})\cdot[\bm{t}(\bm{t}\cdot\bm{u}')+\bm{n}(\bm{n}\cdot\bm{u}')]\text{d}S.
\end{equation}
In our auxiliary problem we have a clean spherical drop which we know does not suffer the effects of any unbalanced tangential stress. Hence $\bm{n}\cdot(\bm{\sigma}_{aux}-\beta\hat{\bm{\sigma}}_{aux})\cdot\bm{t}=0$ for a clean spherical drop. We apply boundary condition mentioned in (\ref{3.3b}), $\bm{n\cdot u}'=-\bm{n}\cdot \bm{u}^\infty$ on the normal component of the integrand
\begin{equation}\label{3.10}
  \int_S[(\beta -1)\bm{n}\cdot \bm{\sigma}^\infty\cdot (\bm{I}-\bm{nn})+Ma\nabla_s\Gamma]\cdot\bm{u}_{aux}\text{d}S = - \int_S\bm{n}\cdot(\bm{\sigma}_{aux}-\beta\hat{\bm{\sigma}}_{aux})\cdot\bm{n}(\bm{n}\cdot \bm{u}^\infty)  .
\end{equation}
The already known solution of the Hadamard-Rybczynski problem (auxiliary problem)(Appendix \ref{appen:Hadamard}) and is substituted in $(\ref{3.10})$
\begin{equation}\label{3.11}
    \int_S[(\beta -1)\bm{n}\cdot \bm{\sigma}^\infty\cdot (\bm{I}-\bm{nn})+Ma\nabla_s\Gamma]\text{d}S=-\int_S 3(3\beta+2)\bm{n}(\bm{n}\cdot \bm{u}^\infty)\text{d}S.
\end{equation}
We then expand the far field flow terms about the origin using Taylor series. The odd terms in the integral such as $\int_S\bm{r}dS$, $\int_S\bm{rrr}dS$ and so on, vanishes due to the symmetry of the sphere. Such terms will be an odd function having equal contributions from positive and negative positions thus cancelling each other.The even terms involving even order derivatives of order greater than $2$ of far field velocity of the structure $\nabla^{4}\bm{u}^\infty(0)$, $\nabla^{6}\bm{u}^\infty(0)$ and so on are identically zero in the creeping flow regime satisfying the biharmonic requirements of Stokes equation. So compiling the terms remaining after elimination \citep{nadim1990thermocapillary} we get
\begin{equation}\label{3.12}
    0 = \bm{u}^\infty(0)+\frac{\beta}{2(3\beta+2)}\nabla^2\bm{u}^\infty(0)+\frac{Ma}{4\pi(3\beta+2)}\int_S\nabla_s\Gamma \text{d}S.
\end{equation}
Substituting $(\ref{2.9})$, $\bm{u}^{\infty}=\bm{v}^{\infty}-\bm{U}$, in $(\ref{3.12})$ we can find the general analytical solution of the drop velocity as
\begin{equation}\label{3.13}
     \bm{U} = \bm{v}^\infty(0)+\frac{\beta}{2(3\beta+2)}\nabla^2\bm{v}^\infty(0)+\frac{Ma}{4\pi(3\beta+2)}\int_S\nabla_s\Gamma \text{d}S.
\end{equation}

\subsection{The spherical harmonics representation of surfactant concentration}\label{sec:sph}
We can use spherical harmonics to define surfactant concentration $(\Gamma)$ as done in \citep{Hanna2010}. Surfactant concentration being a real quantity should be represented using the real basis of spherical harmonics
\begin{equation}\label{3.14}
    \Gamma=1+\sum_{n=1}^\infty\sum_{m=-n}^n g_{n,m}Y_n^m(\theta,\phi).
\end{equation}
Here, $Y_n^m$ are the different modes of spherical harmonics for different values of $n,m$ and $g_{n,m}$ are their coefficients. $Y_n^m$ is defined as
\begin{equation}\label{3.15}
    Y_n^m(\theta,\phi) = \begin{cases}\sqrt{2}k_n^mP_n^m(\cos\text{ }\theta)\text{ }\cos\text{ }m\phi &\quad\text{   if  }m>0,\\
    k_n^0P_n^0(\cos\text{ }\theta) &\quad\text{  if  } m=0,\\
    \sqrt{2}k_n^{|m|}P_n^{|m|}(\cos\text{ }\theta)\text{ }\sin\text{ }|m|\phi &\quad\text{   if  }m<0.\end{cases}
\end{equation}
where $P_n^m$ is the associated Legendre polynomial of order m and degree n and $k_n^m$ is the normalization constant
\begin{equation}
    k_n^m = \sqrt{\frac{(2n+1)(n-m)!}{4\pi(n+m)!}}
\end{equation}
By performing the integration $\int_S\nabla_s\Gamma dS$ , it is seen that only three modes of $(\ref{3.14})$ have notable contribution to the integral
\begin{equation}\label{3.17}
    \int_S\nabla_s\Gamma dS=4\sqrt{\frac{\pi}{3}}(-g_{1,1}\bm{e}_x-g_{1,-1}\bm{e}_y+g_{1,0}\bm{e}_z).
\end{equation}
Substituting $(\ref{3.17})$ in $(\ref{3.13})$ we get
\begin{equation}\label{3.18}
    \bm{U} = \bm{v}^\infty(0)+\frac{\beta}{2(3\beta+2)}\nabla^2\bm{v}^\infty(0)+\frac{Ma}{\sqrt{3\pi}(3\beta+2)}(-g_{1,1}\bm{e}_x-g_{1,-1}\bm{e}_y+g_{1,0}\bm{e}_z).
\end{equation}
\\
The three modes of surfactant concentration distribution propels the migration of the drop. Surfactant distribution for the mode $g_{1,0}$ can be visualized as being concentrated towards one pole of the spherical drop along $z$ axis with respect to the opposite pole. Such distribution of surfactant creates an interfacial tension gradient which propels the drop in $z$ direction. Modes $g_{1,1}$ and $g_{1,-1}$ exhibit similar surfactant distributions along $x$ and $y$ axis respectively thereby propelling the drops along these directions. The coefficients $g_{n,m}$ are determined by projecting the surface transport equation onto spherical harmonics and exploiting the orthogonal property of the spherical harmonics.
\\
For further analysis we define the velocity of an undisturbed unbounded cylindrical Poiseuille flow as the background flow of the drop. Here the flow is expressed with respect to the center of the drop and made dimensionless by scaling it against the characteristic velocity scale $U_b$
\begin{equation}\label{3.19}
    \bm{v}^\infty=\left[1-\left(\frac{r}{R_0}\right)^2\sin^2\theta-\left(\frac{b}{R_0}\right)^2-\left(\frac{2rb}{R_0^2}\right)\sin\theta\cos\phi\right]\bm{e}_z.
\end{equation}

\section{Droplet migration for $Pe_{s}\ll1$}\label{sec:Surfactant induced migration}
\subsection{Regular perturbation expansion in $Pe_s$}\label{sec:regpert}
We analyse the problem defined by $(\ref{2.8})$-$(\ref{2.11})$ perturbatively in the limit of low surface P\'eclet number using the expression $(\ref{3.18})$ developed in the previous section with the help of the reciprocal theorem. We expand the velocity, pressure fields, surfactant concentration and migration velocity in the low surface  P\'eclet number limits using regular perturbation, as,
\begin{equation}\label{4.1}
    [\bm{u},\hat{\bm{u}},p,\hat{p},\Gamma,\bm{U}]=[\bm{u}_0,\hat{\bm{u}}_0,p_0,\hat{p}_0,\Gamma_0,\bm{U}_0]+Pe_{s}[\bm{u}_1,\hat{\bm{u}}_1,p_1,\hat{p}_1,\Gamma_1,\bm{U}_1]+\mathcal{O}({Pe_{s}}^2).
\end{equation}
The equation for surfactant transport as derived in $(\ref{2.11})$ is then expanded following $(\ref{4.1})$ and then collecting the like order terms we obtain the following zeroth-order, first-order and second-order surfactant transport equations
\begin{subequations}
\begin{align}
    \mathcal{O}(1):\quad {\nabla_{s}^2}\Gamma_{0}&=k\left(\Gamma_{0}-1\right),\label{4.2a} \\
    \mathcal{O}(Pe_s): \quad {\nabla_{s}^2}\Gamma_{1}&=\nabla_{s}\cdot\left(\Gamma_{0}\bm{u_{s0}}\right)+k\Gamma_{1},\label{4.2b} \\
    \mathcal{O}({Pe_{s}^2}): \quad {\nabla_{s}^2}\Gamma_{2}&=\nabla_{s}\cdot\left(\Gamma_{1}\bm{u_{s0}}+\Gamma_{0}\bm{u_{s1}}\right)+k\Gamma_{2},\label{4.2c}
\end{align}
\end{subequations}
where $\Gamma_{1}$ and $\Gamma_{2}$ represents the first-order and second-order surfactant concentrations. In addition $\bm{u}_{os}$ and $\bm{u}_{1s}$ are the zeroth-order and first-order tangential velocity components on the drop surface.
\subsection{Zeroth-order solution}\label{sec:zerocorr}
The surfactant transport equation for zeroth-order is given by (\ref{4.2a}).
Because there is no surfactant concentration gradient on a clean spherical drop surface, we have $\nabla_{s}\Gamma_{0}=0$ and $k\neq0$. Hence, \ref{4.2a} will simply boil down to a uniform surfactant concentration $\Gamma_{0}=1$. The velocity fields at this order $(\bm{u}_0)$ are identical to the solution provided by \citep{Hetsroni1970} and \citep{Nadim1991}. At zeroth-order, there is no surfactant-induced migration, hence the equation simplifies to Faxen's law. As expected, there is no surfactant-induced migration at zeroth-order and it reduces to that given by Fax\'en's law $(\ref{3.13})$. For our study the non-dimensional migration velocity at zeroth-order in an unbounded Poiseuille flow (\ref{3.19}), is given by
\begin{equation}\label{4.3}
    \bm{U}_0=\bm{v}^\infty(0)+\frac{\beta}{2(3\beta+2)}\nabla^2\bm{v}^\infty(0)=\left[1-\left(\frac{b}{R_0}\right)^2-\frac{2\beta}{(3\beta+2){R_{0}^2}}\right]\bm{e}_z.
\end{equation}
\subsection{First-order correction}\label{sec:firstorder}
The first-order corrected velocity fields and pressure are obtained by substituting $(\ref{4.1})$ in $(\ref{2.8})$. It can be shown that they will satisfy Stokes and continuity equations and are given by
\begin{subequations}\label{4.4}
\begin{equation}
     \nabla p_1=\nabla^2 \bm{u}_1, \quad \nabla \cdot \bm{u}_1=0,
\end{equation}
\begin{equation}
     \nabla \hat{p_1}=\nabla^2 \hat{\bm{u}}_1, \quad \nabla \cdot \hat{\bm{u}}_1=0.
\end{equation}
\end{subequations}
The far-field velocity approaches the uniform flow
\begin{equation}
    \bm{u}_1\left(r\rightarrow\infty\right)=-\bm{U}_1,
\end{equation}
where $\bm{U}_1$ is the unknown first-order corrected migration velocity. The kinematic boundary conditions at the interface $(\ref{4.1})$ may be obtained using $(\ref{4.1})$ as follows:
\begin{subequations}
\begin{align}
    \bm{u}_1&=\hat{\bm{u}}_1,\\
    \bm{n}\cdot\bm{u}_1&=0.
\end{align}
\end{subequations}
and using $(\ref{4.1})$, the tangential stress balance is given by
\begin{equation}\label{4.7}
    \bm{n}\cdot\left(\sigma_{1}-\beta\hat{\sigma}_1\right)\cdot\left(\bm{I}-\bm{nn}\right)=Ma\nabla_{s}\Gamma_{1}.
\end{equation}
Since $\Gamma_{0}=1$ the first-order transport equation given by (\ref{4.2b}) can be reduced to 
\begin{equation}
    {\nabla_{s}^2}\Gamma_{1}=\nabla_{s}\cdot\bm{u_{s0}}+k\Gamma_{1}.
\end{equation}
We express $\Gamma_{1}$ in terms of spherical harmonics to get the first-order surfactant concentration $\Gamma_{1}$,
\begin{equation}
    \Gamma_{1}=\sum_{n=1}^{\infty}\sum_{m=-n}^{n}{g_{n,m}^{(1)}}{Y_{n}^{m}}.
\end{equation}
Using the property that ${\nabla_{s}^2}\Gamma_{1}=-\sum_{n=1}^{\infty}\sum_{m=-n}^{n} n(n+1){g_{n,m}^{(1)}}{Y_{n}^{m}}$ and by projecting $\nabla_{s}\cdot\bm{u_{s0}}$ onto the basis of spherical harmonics, the coefficients ${g_{n,m}^{(1)}}$ are easily determined as,
\begin{equation}\label{4.10}
    {g_{n,m}^{(1)}}=-\frac{1}{k+n(n+1)}\int_0^{2\pi}\int_0^\pi\left(\nabla_{s}\cdot\bm{u_{s0}}\right){Y_{n}^{m}}\sin{\theta}\text{d}\theta \text{d}\phi.
\end{equation}
The zeroth-order tangential component of velocity on the surface of the drop, $u_{s0}$, in the unbounded background Poiseuille flow is calculated following the works of \citet{Hetsroni1970} and \citet{Nadim1991}. The first-order surfactant concentration is calculated using projections as
\begin{align}\label{eqn:4.11 gamma }
     \Gamma_{1}(\theta,\phi)=&\frac{1}{R_{0}^2}[-\frac{4}{(k+2)(2+3\beta)}{P_{1}^{0}}(\cos{\theta})+\frac{2}{(k+12)(1+\beta)}{P_{3}^{0}}(\cos{\theta})\nonumber\\
       &+\frac{2b}{(k+6)(1+\beta)}{P_{2}^{1}}(\cos{\theta})\cos{\phi}].
\end{align}
The full calculation of $\Gamma_{1}$ was not required but was done to make the calculation look thorough and complete, as from the derived result of the reciprocal theorem $(\ref{3.18})$ we need to calculate only three projections for the coefficients $ {g_{1,1}^{(1)}}$, $ {g_{1,-1}^{(1)}}$ and $ {g_{1,0}^{(1)}}$ in order to compute the migration velocity at this order induced by the surfactant concentration,
\begin{equation}\label{4.12}
    \bm{U}_{1}=\frac{Ma}{4\pi(3\beta+2)}\int_{S}\nabla_{s}\Gamma_{1}\text{d}S=\frac{Ma}{\sqrt{3\pi}(3\beta+2)}\left(-{g_{1,1}^{(1)}}\bm{e}_{x}-{g_{1,-1}^{(1)}}\bm{e}_{y}+{g_{1,0}^{(1)}}\bm{e}_{z}\right).
\end{equation}
Using the reciprocal theorem approach, we have avoided the intricate and involved computation required for solving the first-order problem $(\bm{u}_{1}$ and $p_1)$. 
\\ 
Carrying out the three essential projections in $(\ref{4.10})$, we get
\begin{subequations}
\begin{equation}
    {g_{1,1}^{(1)}}=0,
\end{equation}
\begin{equation}
    {g_{1,-1}^{(1)}}=0,
\end{equation}
\begin{equation}
    {g_{1,0}^{(1)}}=-8\sqrt{\frac{\pi}{3}}\frac{1}{(k+2)(3\beta+2){R_{0}^2}}.
\end{equation}
\end{subequations}
As result, by inserting these values in $(\ref{4.12})$, we obtain the first-order migration velocity of a neutrally buoyant spherical drop in an unbounded cylindrical Poiseuille flow as
\begin{equation}\label{4.14}
    \bm{U}_{1}=-\frac{8Ma}{3(k+2){(3\beta+2)}^2{R_{0}^2}}\bm{e}_z.
\end{equation}
It's worth noting that there is no velocity component in the $x$ and $y$ directions at this order. When compared to the dynamics for a clean interface, $(\ref{4.3})$, this may be considered as a slip velocity of the drop in the flow direction induced by the leading-order effect of surfactant redistribution and reaction, resulting in a slowed down speed of the drop. This slip velocity $(\ref{4.14})$ does not depend on $D$, the position of the drop in Poiseuille flow but is dependent on $k$, the 1\textsuperscript{st} order dimensionless rate constant. For $k=0$ (\ref{4.14}) reduces down to the same qualitative result as presented by \citet{pak2014viscous} for the first-order surfactant-induced migration velocity correction and has an identical characteristic as the lowered rise speed of a buoyant drop \citep{levich1962physicochemical}.
\subsection{Second-order correction}\label{sec:seccorr}
Proceeding in an identical manner as we did in the previous section we can derive a similar set of equations as given in $(\ref{4.4})$-$(\ref{4.7})$.\\
The second-order surfactant transport equation (\ref{4.2c}), determines surfactant concentration at this order. Since, $\Gamma_{0}=1$, the revised transport equation becomes,
\begin{equation}
    {\nabla_{s}^2}\Gamma_{2}=\nabla_{s}\cdot\left(\Gamma_{1}\bm{u_{s0}}+\bm{u_{s1}}\right)+k\Gamma_{2}
\end{equation}
We express $\Gamma_{2}$ in terms of spherical harmonics in an analogous manner as,
\begin{equation}
    \Gamma_{2}=\sum_{n=1}^{\infty}\sum_{m=-n}^{n}{g_{n,m}^{(2)}}{Y_{n}^{m}}.
\end{equation}
The coefficients ${g_{n,m}^{(2)}}$ can be calculated as
\begin{equation}
    {g_{n,m}^{(2)}}=-\frac{1}{k+n(n+1)}\int_0^{2\pi}\int_0^\pi\nabla_{s}\cdot\left(\Gamma_{1}\bm{u_{s0}}+\bm{u_{s1}}\right){Y_{n}^{m}}\sin{\theta}\text{d}\theta \text{d}\phi.
\end{equation}
But in order to compute the coefficients ${g_{n,m}^{(2)}}$ we need to solve the first-order problem defined in $(\ref{4.4})$-$(\ref{4.7})$ for the velocity field $\bm{u}_1$ and $\hat{\bm{u}}_1$ in order to determine $\bm{u_{s1}}$. After determining $\bm{U_{1}}$ in $(\ref{4.14})$, the first-order problem is then worked out by referring to \citet{Palaniappan1992} for the general velocity and pressure representations in Stokes flows, which is identical to  Lamb’s general solution.
The first-order tangential velocity on the droplet surface, $\bm{u_{s1}}$ in the Poiseuille flow is calculated using the results of \citet{Choudhuri2013}, who carried out a similar kind of methodology for studying the thermocapillary motion of a viscous drop.\\
With the aid of the result from the reciprocal theorem $(\ref{3.13})$, the surfactant-induced migration velocity at this order is given by,
\begin{equation}\label{4.18}
    \bm{U}_{2}=\frac{Ma}{4\pi(3\beta+2)}\int_{S}\nabla_{s}\Gamma_{2}\text{d}S=\frac{Ma}{\sqrt{3\pi}(3\beta+2)}\left(-{g_{1,1}^{(2)}}\bm{e}_{x}-{g_{1,-1}^{(2)}}\bm{e}_{y}+{g_{1,0}^{(2)}}\bm{e}_{z}\right).
\end{equation}
Therefore the three necessary coefficients contributing to the migration velocity at this order are obtained as,
\begin{subequations}
\begin{equation}
    {g_{1,1}^{(2)}}=-\frac{8}{35}\sqrt{\frac{\pi}{3}}\frac{b\left(35k^2(1+\beta)^2+36(40+109\beta+70\beta^2)+6k(85+187\beta+105\beta^2)\right)}{(k+2)^2(k+6)(k+12)(1+\beta)^2(3\beta+2){R_{0}^2}},
\end{equation}
\begin{equation}
    {g_{1,-1}^{(2)}}=0,
\end{equation}
\begin{equation}
    {g_{1,0}^{(2)}}=8\sqrt{\frac{\pi}{3}}\frac{1}{(k+2)(3\beta+2)^2{R_{0}^2}}.
\end{equation}
\end{subequations}
Therefore, using $(\ref{4.18})$ the second-order surfactant-induced migration velocity of the drop is,
\begin{align}\label{4.20}
    \bm{U}_2=&-\frac{8bMa}{105}\frac{\left[35k^2(1+\beta)^2+36(40+109\beta+70\beta^2)+6k(85+187\beta+105\beta^2)\right]}{(k+2)^2(k+6)(k+12)(1+\beta)^2(3\beta+2){R_{0}^2}}\bm{e}_{x}\nonumber\\&+\frac{8Ma^2}{3(k+2)(3\beta+2)^2{R_{0}^4}}\bm{e}_{z}.
\end{align}
It is important to observe that even at this order there is no migration velocity in the $y$ direction due to symmetry. But interestingly, a cross-streamline component along the $x$ direction appears at this order which is transverse to the flow field. The magnitude of the cross-streamline velocity is linearly dependent on $D$, the distance of the drop from the centre of the Poiseuille flow, but is dependent on the dimensionless rate constant, $k$, in a highly non-linear fashion. We also notice a further correction to the slip velocity in the $z$ direction, which, in contrary to that of first order, increases the drop velocity in $z$ direction. The slip velocity at this order is again found to be independent of $D$. For $k=0$  (\ref{4.20}) reduces down to the same qualitative nature as presented by \citet{pak2014viscous} for the second-order  migration velocity of the drop.
\section{Results and discussion}\label{sec:results}
In our study, we demonstrate a general analysis of the motion of an undeformed spherical drop in an unbounded cylindrical Poiseuille flow with reacting solute at the droplet-suspending fluid interface. \\
The migration velocity of a force-free drop presented in our study is
\begin{align}\label{5.1}
      \bm{U}=&\bm{U}_{0}+Pe_{s}\bm{U}_{1}+{Pe_{s}^2}\bm{U}_{2}+\mathcal{O}({Pe_{s}^3})\nonumber\\
    =&\left[1-\left(\frac{b}{R_0}\right)^2-\frac{2\beta}{(3\beta+2){R_{0}^2}}-Pe_{s}\frac{8Ma}{3(k+2)(3\beta+2)^2{R_{0}^2}}\left(1-Pe_{s}\frac{Ma}{3\beta+2}\right)\right]\bm{e}_{z}\nonumber\\
    &-{Pe_{s}^2}\frac{8bMa}{105}\frac{\left[(35k^2(1+\beta)^2+36(40+109\beta+70\beta^2)+6k(85+187\beta+105\beta^2)\right]}{(k+2)^2(k+6)(k+12)(1+\beta)^2(3\beta+2){R_{0}^4}}\bm{e}_{x}+\mathcal{O}({Pe_{s}^3}).
\end{align}
\\
In zeroth order, the outcomes are identical to a clean drop in an unbounded Poiseuille flow.. The zeroth order velocity in its dimensional form $\bm{U}_0^*$ is found to be linearly dependent on our velocity scale $\bm{U}_b$, i.e., $\bm{U}_0^*=\bm{U}_b\bm{U}_0\propto\bm{U}_b $. This ensures there will be a change in direction of the migration velocity if there is a change in direction of the background flow.

\par
The first order migration velocity exhibits a similar pattern. It is found to be directly proportional to the background flow velocity, $\bm{U}_1^*\propto Pe_sMa\bm{U}_b$. This satisfies the symmetry requirement ensuring change in direction of migration velocity with change in direction of background flow. It is also seen that there is no migration in the transverse direction at first order. There is no term in the $\bm{e}_x$ direction, i.e., $\bm{U}_{1x}=0$.
\par
Analysing the second order terms of migration velocity the $z$ direction term is again found to be linearly dependent on the background flow similar to the other orders i.e., $\bm{U}_{2z}^*\propto Pe_s^2 Ma^2 \bm{U}_b\propto\bm{U}_b$. On the other hand the dimensional cross streamline migration velocity is found to exhibit a quadratic dependence on the background flow ie,\\ $\bm{U}_{2x}^*\propto Pe_s^2 Ma \bm{U}_b\propto\bm{U}_b^2$. This is because the product of $Pe_s^2$ and $Ma$ is directly proportional to $\bm{U}_b$.  The quadratic dependence indicate that there will be no change in direction of migration in transverse direction with change in direction of background flow. It is evident from $(\ref{5.1})$, that the drop always migrates towards the center of the Poiseuille flow, regardless of the direction of the background flow. Moreover, the amplitude of the drop is linearly proportional to its distance from the Poiseuille flow's centerline, $D$. $\bm{U}_{2x}$ being directly proportional to $D$ the symmetry condition is maintained since the transverse migration changes direction when the drop is placed on the opposite side with respect to the centerline of the Poiseuille flow, i.e., $b\to -b$.
\par
We have considered quasi-steady state approximation to find an analytical solution in this problem. The drop when placed in the Poiseuille flow encounters a continuous change in background flow, velocity and surfactant distribution depending on its position in the background flow until it reaches the center line. As a result it can never achieve steady state at any position other than the center of the Poiseuille flow as it needs to continuously adapt to its surroundings and undergo transverse migration. So to find an analytical solution we consider quasi steady state approximation owing to the fact that the time scale for diffusion and background flow is much lesser than the time scale for longitudinal and transverse migration velocity of the drop. The velocity of the background flow shows instantaneous effect in the Stokes flow regime. It is safe to infer that the change in surfactant distribution and establishing of velocity field is adequately fast with respect to the speed of the drop thereby validating our assumption.
\par
\begin{figure}
\centering
    \subfloat[]{\includegraphics[width=0.5\textwidth]{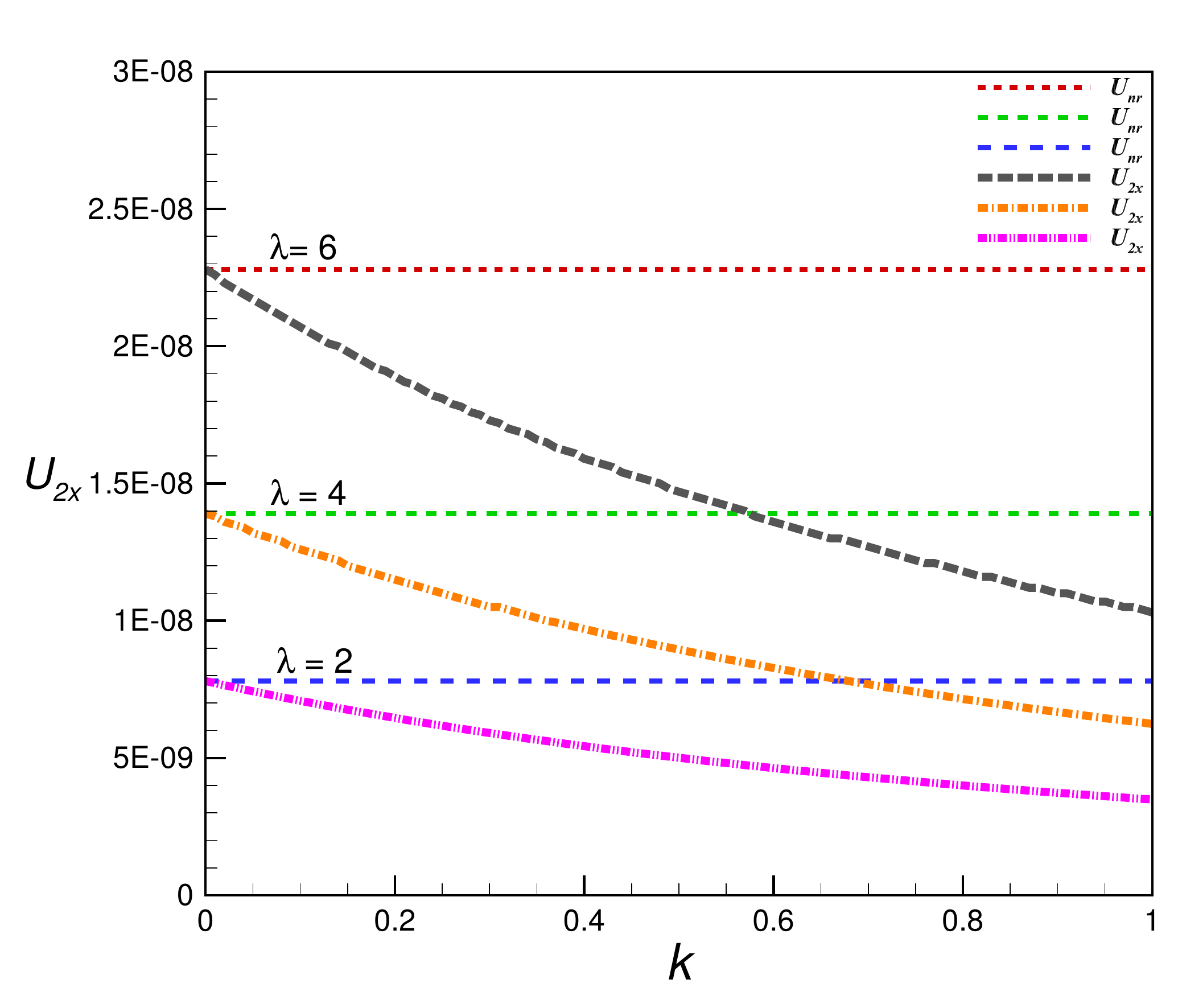}\label{fig:Ux vs k lambda >1}}
     \subfloat[]{\includegraphics[width=0.5\textwidth]{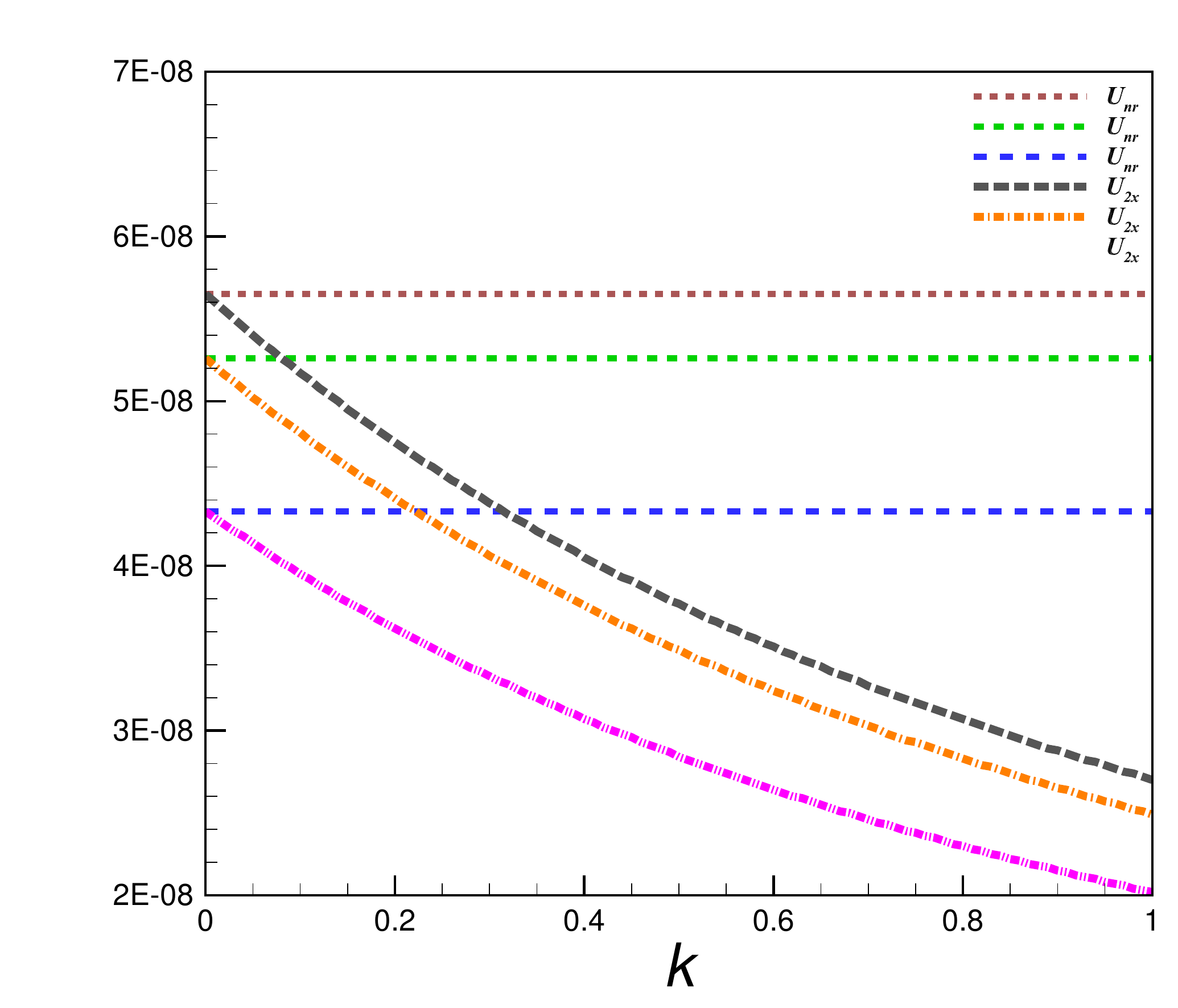}\label{fig:Ux vs k lambda <1}}
     \caption{(a) Variation of cross-stream migration velocity for both reacting $(U_{2x})$ and non-reacting solute $(U_{nr})$ with dimensionless rate constant $k$ for $\beta>1$ (b) Variation of cross-stream migration velocity for both reacting $(U_{2x})$ and non-reacting solute $(U_{nr})$ with dimensionless rate constant $k$ for $\beta<1$ .}
\end{figure}
\subsection{Effect of surface reaction}\label{sec:effreact}
As the results obtained in zeroth order resemble to that of a clean drop in an unbounded Poiseuille flow, no added effect due to reaction is observed in the leading order.
\par
An effect of reaction is observed in the first order drop migration velocity. The dimensionless reaction rate constant $k$ is always greater than zero and it is found in the denominator of the first order term of (\ref{5.1}). Therefore surface reaction is causing a retarding effect in the first order drop migration velocity in $z$ direction. Comparing the first order term of $(\ref{5.1})$ with the results obtained by \citet{pak2014viscous} it can be confirmed that longitudinal drop migration would have been faster in absence of surface reaction. No additional effect on transverse migration due to surface reaction is observed for first order.
\par
Effect of surface reaction on second order terms is found to be similar to that of first order. The reaction rate constant being in denominator has identical impeding effect on longitudinal drop migration velocity as can be seen when compared with the results of \citet{pak2014viscous}. The cross streamline migration velocity in the second order term also experiences a slowed down effect due to surface reaction. This can be observed by plotting and comparing the results obtained by \citet{pak2014viscous} which is devoid of effect of reaction with the term in $\bm{e}_x$ direction in (\ref{5.1}). Therefore it can be inferred that both longitudinal and transverse drop migration velocity would have been higher in absence of surface reaction. 

\par
The validity of our results can be confirmed by substituting $k=0$ in (\ref{5.1}) which gives us the migration velocity obtained by \citet{pak2014viscous} exactly. The overall migration obtained by \citet{pak2014viscous} was :
\begin{align}
      \bm{U}=&\bm{U}_{0}+Pe_{s}\bm{U}_{1}+{Pe_{s}^2}\bm{U}_{2}+\mathcal{O}({Pe_{s}^3})\nonumber\\
    =&\left[1-\left(\frac{b}{R_0}\right)^2-\frac{2\beta}{(3\beta+2){R_{0}^2}}-Pe_{s}\frac{4Ma}{3(3\beta+2)^2{R_{0}^2}}\left(1-Pe_{s}\frac{Ma}{3\beta+2}\right)\right]\bm{e}_{z}\nonumber\\
   &-{Pe_{s}^2}\frac{bMa}{105}\frac{(40+109\beta+70\beta^2)}{(1+\beta)^2(3\beta+2){R_{0}^4}}\bm{e}_{x}+\mathcal{O}({Pe_{s}^3}).
\end{align}
\par
The analytical model developed in this study is simulated by
using the following choices of system parameters for the
sake of illustration: $Ma=1$, $Pe_s=0.01$, $R_0=5$ and $b=2$.
To find out how important the dimensionless rate constant $k$ is for the drop's cross-stream migration velocity, we plot and compare the variation of $U_x$ with $k$ for reacting and non-reacting solute ($k=0$) . It is evident from Fig. \ref{fig:Ux vs k lambda >1} that with increasing $k$ the magnitude of the $x$ component of reactive migration 
velocity decreases gradually and the magnitude of the reactive $U_x$ is always lesser than 
the non-reactive $U_x$. It is observed that for $\beta>1$, with increase in $\beta$, the absolute values of the reactive $U_x$ increases. Fig. \ref{fig:Ux vs k lambda <1} also depicts similar kind of variation of $U_x$ with $k$, but, for $\beta<1$, with increase in $\beta$, the absolute values of the reactive $U_x$ decreases. \par
 \section{Conclusion}
 A theoretical model has been established in this paper for studying the dynamics of a buoyant drop subjected to a cylindrical Poiseuille in the background with a reacting solute at the drop's contact with the suspending fluid. The migration velocity profile has been obtained analytically under the limit of low P\'eclet number using Lorentz reciprocal theorem and regular perturbation analysis. It has been revealed that dimensionless rate constant $k$ significantly decreases the cross-stream migration velocity. 
 \par
 The current research indicates a number of potential directions. Our method, for example, may be used to non-spherical or multiple droplets. In addition how does the dynamics depends on the fluid rheology (thinning vs thickening) is also a question which needs to be addressed. However, we shall save these fascinating topics for future research.

\appendix
\section{The Hadamard–Rybczynski problem}\label{appen:Hadamard}
The solution to the classic Hadamard–Rybczynski problem \citep{leal2007advanced}, which is the auxiliary problem for $\left(\bm{u}_{aux},\bm{\sigma}_{aux}\right)$ and $(\hat{\bm{u}}_{aux},\hat{\bm{\sigma}}_{aux})$ in § \ref{sec:recthrm} is replicated here and the existing results are used in (\ref{3.10}) at the interface $(r=1)$:
\begin{subequations}
\begin{align}
     \bm{u}_{aux}=\hat{\bm{u}}_{aux}=&\frac{1}{2(1+\beta)}(\bm{I}-\bm{nn})\cdot\bm{U}_{aux}\\
     \bm{n}\cdot\bm{\sigma}_{aux}=&\frac{3}{2(1+\beta)}(\beta\bm{I}+2\bm{nn})\cdot\bm{U}_{aux}\\
     \bm{n}\cdot(\bm{\sigma}_{aux}-\beta\hat{\bm{\sigma}}_{aux})\cdot\bm{n}=&\frac{3(2+3\beta)}{2(1+\beta)}\bm{n}\cdot\bm{U}_{aux}
\end{align}
\end{subequations}
\section{The zeroth-order solution}\label{appen:Zeroth order soln}
For a spherical drop immersed in an unbounded cylindrical Poiseuille flow, using a coordinate system centred at the drop's centre, the zeroth-order solution in § \ref{sec:firstorder} may be computed using \citet{Hetsroni1970} and \citet{Nadim1991}'s work. The zeroth-order velocity field outside
the drop $(\bm{u}_0=u_{0r}\bm{e}_r+u_{0\theta}\bm{e}_{\theta}+u_{0\phi}\bm{e}_{\phi})$ is given by
\begin{subequations}
\begin{flalign}
    u_{0r} = &\frac{\cos{\theta}}{2r^5(1+\beta)(2+3\beta){R^2}_0}(4r^5\beta(1+\beta)-3\beta(2+3\beta)+r^2(4+12\beta+11\beta^2) \nonumber\\&-(2+3\beta)[(-5\beta+2r^2+7\beta r^2)\cos^2{\theta}+2r\sin{\theta}(3b\beta\cos{\phi}-2r^2b\cos{\phi}\nonumber\\&-5r^2b\beta\cos{\phi}+2r^5b\cos{\phi}+2r^5b\beta\cos{\phi}+r^6\sin{\theta}+r^6\beta\sin{\theta})]),&&
\end{flalign}
\begin{flalign}
    u_{0\theta}=&\frac{1}{32r^5(1+\beta)(2+3\beta){R^2}_0}(32br\beta(2+3\beta)\cos^2{\theta}\cos{\phi}+\sin{\theta}[-64r^5\beta(1+\beta)\nonumber\\&+3\beta(2+3\beta)+r^2(12-4\beta-17\beta^2)+32r(2+3\beta)\sin{\theta}(-b\beta\cos{\phi}\\&+2r^5b\cos{\phi}+2r^5b\beta\cos{\phi}+r^6\sin{\theta}+r^6\beta\sin{\theta})]-(2+3\beta)(-15\beta+2r^2+7\beta r^2)\sin{3\theta}),\nonumber&&
\end{flalign}
\begin{flalign}
u_{0\phi}=-\frac{b\beta\cos{\theta}\sin{\phi}}{r^4(1+\beta){R^2}_0}, &&
\end{flalign}
\end{subequations}
and the velocity field inside the drop $(\hat{\bm{u}_0}=\hat{u}_{0r}\bm{e}_r+\hat{u}_{0\theta}\bm{e}_{\theta}+\hat{u}_{0\phi}\bm{e}_{\phi})$ is given by
\begin{subequations}
\begin{flalign}
    \hat{u}_{0r}=&\frac{r^2-1}{4(1+\beta)(2+3\beta){R^2}_0}(\cos{\theta}[-8(1+\beta)-r^2(2+3\beta)+5r^2(2+3\beta)\cos{2\theta}]\nonumber\\&-6br(2+3\beta)\cos{\phi}\sin{2\theta}), &&
\end{flalign}
\begin{flalign}
    \hat{u}_{0\theta}=&\frac{1}{4(1+\beta)(2+3\beta){R^2}_0}(-2br(2+3\beta)[-2(1+\beta)+(-3+5r^2)\cos{2\theta}]\cos{\phi}\nonumber\\&+[-8(1+\beta)+2r^2(6+5\beta)+r^4(6+9\beta)-5r^2(-2+3r^2)(2+3\beta)\cos^2{\theta}]\sin{\theta}), &&
\end{flalign}
\begin{flalign}
     \hat{u}_{0\phi}=\frac{br(5r^2-2\beta-5)\cos{\theta}\sin{\phi}}{2(1+\beta){R^2}_0}. &&
\end{flalign}
\end{subequations}
The pressure fields outside $(p_0)$ and inside $(\hat{p}_0)$ the drop are given, respectively, by
\begin{subequations}
\begin{flalign}
    p_0= \frac{\cos{\theta}}{4r^4(1+\beta){R^2}_0}(6+21\beta-16r^5(1+\beta)-5(2+7\beta)\cos^2{\theta}+8br(2+5\beta)\cos{\phi}\sin{\theta}), &&
\end{flalign}
\begin{flalign}
    \hat{p}_0=\frac{r\cos{\theta}}{(1+\beta)(2+3\beta){R^2}_0}[-20(1+\beta)-9r^2(2+3\beta)+3r(2+3\beta)(5r\cos^2{\theta}-7b\cos{\phi}\sin{\theta}]. &&
\end{flalign}
\end{subequations}
\section{The first-order solution}\label{appen:1st order soln}
The first-order solution in  § \ref{sec:seccorr} is obtained from the works of \citet{Choudhuri2013} and \citet{Palaniappan1992} using representation for the general velocity and pressure fields in Stokes flows. The velocity outside the drop $(\bm{u}_1=u_{1r}\bm{e}_r+u_{1\theta}\bm{e}_{\theta}+u_{1\phi}\bm{e}_{\phi})$ in spherical coordinates is given by
\begin{subequations}
\begin{flalign}
    u_{1r}=&\frac{Ma(r-1)\cos{\theta}}{420r^5(1+\beta)^2(2+3\beta)^2{R^2}_0}\nonumber\\ &\times[560r^2(1+\beta)^2+560r^3(1+\beta)^2+560r^4(1+\beta)^2+15(2+3\beta)^2\nonumber\\&+15r(2+3\beta)^2-3(2+3\beta)^2(1+r)(25\cos{2\theta}-84br\cos{\phi}\sin{\theta})], &&
\end{flalign}
\begin{flalign}
    u_{1\theta}=&\frac{-5Ma}{1680r^5(1+\beta)^2(2+3\beta)^2{R^2}_0}\nonumber\\ &\times([448r^5(1+\beta)^2-27(2+3\beta)^2+r^2(260+556\beta+305\beta^2)]\sin{\theta}\nonumber\\&+3(2+3\beta)^2[112br\cos{\phi}-25(r^2-3)\sin{\theta}]\cos{2\theta}), &&
\end{flalign}
\begin{flalign}
    u_{1\phi}=-\frac{bMa\cos{\theta}\sin{\phi}}{5r^4(1+\beta)^2{R^2}_0}, &&
\end{flalign}
\end{subequations}
and the first-order velocity inside the drop $(\hat{\bm{u}_1}=\hat{u}_{1r}\bm{e}_r+\hat{u}_{1\theta}\bm{e}_{\theta}+\hat{u}_{1\phi}\bm{e}_{\phi})$ is given by
\begin{subequations}
\begin{flalign}
   \hat{u}_{1r}=& \frac{Ma(r^2-1)\cos{\theta}}{140(1+\beta)^2(2+3\beta)^2{R^2}_0}\nonumber\\ &\times[280(1+\beta)^2+5r^2(2+3\beta)^2+r(2+3\beta)^2(-25r\cos{2\theta}+84b\cos{\phi}\sin{\theta}],&&
\end{flalign}
\begin{flalign}
    \hat{u}_{1\theta}=&\frac{Ma}{280(1+\beta)^2(2+3\beta)^2{R^2}_0}\nonumber\\
    &\times([560(1+\beta)^2+45r^4(2+3\beta)^2-10r^2(124+260\beta+139\beta^2)]\sin{\theta}\nonumber\\
    &+r(2+3\beta)^2[28b(-3+5r^2)\cos{\phi}+25r(3r^2-2)\sin{\theta}]\cos{2\theta}),&&
\end{flalign}
\begin{flalign}
      \hat{u}_{1\phi}=\frac{bMar(3-5r^2)\cos{\theta}\sin{\phi}}{10(1+\beta)^2{R^2}_0}. &&
\end{flalign}
\end{subequations}
The first-order pressure fields outside $(p_1)$ and inside $\hat{p}_1$ the drop are given,
respectively, by
\begin{subequations}
\begin{flalign}
    p_1=\frac{Ma(-75\cos{\theta}-125\cos{3\theta}+336br\cos{\phi}\sin{2\theta})}{560r^4(1+\beta)^2{R^2}_0}, &&
\end{flalign} 
\begin{flalign}
     \hat{p}_1=&\frac{Ma r}{140(1+\beta)^2(2+3\beta)^2{R^2}_0}\nonumber\\
&\times([2800(1+\beta)^2-45r^2(2+3\beta)^2]\cos{\theta}
+3r(2+3\beta)^2(-25r\cos{3\theta}+98b\cos{\phi}\sin{2\theta})).&&
\end{flalign}
\end{subequations}

\nocite{*}
\bibliography{aipsamp}

\providecommand{\noopsort}[1]{}\providecommand{\singleletter}[1]{#1}%
\begin{thebibliography}{38}%
\makeatletter
\providecommand \@ifxundefined [1]{%
 \@ifx{#1\undefined}
}%
\providecommand \@ifnum [1]{%
 \ifnum #1\expandafter \@firstoftwo
 \else \expandafter \@secondoftwo
 \fi
}%
\providecommand \@ifx [1]{%
 \ifx #1\expandafter \@firstoftwo
 \else \expandafter \@secondoftwo
 \fi
}%
\providecommand \natexlab [1]{#1}%
\providecommand \enquote  [1]{``#1''}%
\providecommand \bibnamefont  [1]{#1}%
\providecommand \bibfnamefont [1]{#1}%
\providecommand \citenamefont [1]{#1}%
\providecommand \href@noop [0]{\@secondoftwo}%
\providecommand \href [0]{\begingroup \@sanitize@url \@href}%
\providecommand \@href[1]{\@@startlink{#1}\@@href}%
\providecommand \@@href[1]{\endgroup#1\@@endlink}%
\providecommand \@sanitize@url [0]{\catcode `\\12\catcode `\$12\catcode
  `\&12\catcode `\#12\catcode `\^12\catcode `\_12\catcode `\%12\relax}%
\providecommand \@@startlink[1]{}%
\providecommand \@@endlink[0]{}%
\providecommand \url  [0]{\begingroup\@sanitize@url \@url }%
\providecommand \@url [1]{\endgroup\@href {#1}{\urlprefix }}%
\providecommand \urlprefix  [0]{URL }%
\providecommand \Eprint [0]{\href }%
\providecommand \doibase [0]{http://dx.doi.org/}%
\providecommand \selectlanguage [0]{\@gobble}%
\providecommand \bibinfo  [0]{\@secondoftwo}%
\providecommand \bibfield  [0]{\@secondoftwo}%
\providecommand \translation [1]{[#1]}%
\providecommand \BibitemOpen [0]{}%
\providecommand \bibitemStop [0]{}%
\providecommand \bibitemNoStop [0]{.\EOS\space}%
\providecommand \EOS [0]{\spacefactor3000\relax}%
\providecommand \BibitemShut  [1]{\csname bibitem#1\endcsname}%
\let\auto@bib@innerbib\@empty
\bibitem [{\citenamefont {Maass}\ \emph {et~al.}(2016)\citenamefont {Maass},
  \citenamefont {Kr{\"u}ger}, \citenamefont {Herminghaus},\ and\ \citenamefont
  {Bahr}}]{maass2016swimming}%
  \BibitemOpen
  \bibfield  {author} {\bibinfo {author} {\bibfnamefont {C.~C.}\ \bibnamefont
  {Maass}}, \bibinfo {author} {\bibfnamefont {C.}~\bibnamefont {Kr{\"u}ger}},
  \bibinfo {author} {\bibfnamefont {S.}~\bibnamefont {Herminghaus}}, \ and\
  \bibinfo {author} {\bibfnamefont {C.}~\bibnamefont {Bahr}},\ }\bibfield
  {title} {\enquote {\bibinfo {title} {Swimming droplets},}\ }\href@noop {}
  {\bibfield  {journal} {\bibinfo  {journal} {Annual Review of Condensed Matter
  Physics}\ }\textbf {\bibinfo {volume} {7}},\ \bibinfo {pages} {171--193}
  (\bibinfo {year} {2016})}\BibitemShut {NoStop}%
\bibitem [{\citenamefont {Yoshinaga}(2017)}]{yoshinaga2017simple}%
  \BibitemOpen
  \bibfield  {author} {\bibinfo {author} {\bibfnamefont {N.}~\bibnamefont
  {Yoshinaga}},\ }\bibfield  {title} {\enquote {\bibinfo {title} {Simple models
  of self-propelled colloids and liquid drops: from individual motion to
  collective behaviors},}\ }\href@noop {} {\bibfield  {journal} {\bibinfo
  {journal} {Journal of the Physical Society of Japan}\ }\textbf {\bibinfo
  {volume} {86}},\ \bibinfo {pages} {101009} (\bibinfo {year}
  {2017})}\BibitemShut {NoStop}%
\bibitem [{\citenamefont {Jin}, \citenamefont {Kr{\"u}ger},\ and\ \citenamefont
  {Maass}(2017)}]{jin2017chemotaxis}%
  \BibitemOpen
  \bibfield  {author} {\bibinfo {author} {\bibfnamefont {C.}~\bibnamefont
  {Jin}}, \bibinfo {author} {\bibfnamefont {C.}~\bibnamefont {Kr{\"u}ger}}, \
  and\ \bibinfo {author} {\bibfnamefont {C.~C.}\ \bibnamefont {Maass}},\
  }\bibfield  {title} {\enquote {\bibinfo {title} {Chemotaxis and
  autochemotaxis of self-propelling droplet swimmers},}\ }\href@noop {}
  {\bibfield  {journal} {\bibinfo  {journal} {Proceedings of the National
  Academy of Sciences}\ }\textbf {\bibinfo {volume} {114}},\ \bibinfo {pages}
  {5089--5094} (\bibinfo {year} {2017})}\BibitemShut {NoStop}%
\bibitem [{\citenamefont {Schmitt}\ and\ \citenamefont
  {Stark}(2013)}]{schmitt2013swimming}%
  \BibitemOpen
  \bibfield  {author} {\bibinfo {author} {\bibfnamefont {M.}~\bibnamefont
  {Schmitt}}\ and\ \bibinfo {author} {\bibfnamefont {H.}~\bibnamefont
  {Stark}},\ }\bibfield  {title} {\enquote {\bibinfo {title} {Swimming active
  droplet: A theoretical analysis},}\ }\href@noop {} {\bibfield  {journal}
  {\bibinfo  {journal} {EPL (Europhysics Letters)}\ }\textbf {\bibinfo {volume}
  {101}},\ \bibinfo {pages} {44008} (\bibinfo {year} {2013})}\BibitemShut
  {NoStop}%
\bibitem [{\citenamefont {Peddireddy}\ \emph {et~al.}(2012)\citenamefont
  {Peddireddy}, \citenamefont {Kumar}, \citenamefont {Thutupalli},
  \citenamefont {Herminghaus},\ and\ \citenamefont
  {Bahr}}]{peddireddy2012solubilization}%
  \BibitemOpen
  \bibfield  {author} {\bibinfo {author} {\bibfnamefont {K.}~\bibnamefont
  {Peddireddy}}, \bibinfo {author} {\bibfnamefont {P.}~\bibnamefont {Kumar}},
  \bibinfo {author} {\bibfnamefont {S.}~\bibnamefont {Thutupalli}}, \bibinfo
  {author} {\bibfnamefont {S.}~\bibnamefont {Herminghaus}}, \ and\ \bibinfo
  {author} {\bibfnamefont {C.}~\bibnamefont {Bahr}},\ }\bibfield  {title}
  {\enquote {\bibinfo {title} {Solubilization of thermotropic liquid crystal
  compounds in aqueous surfactant solutions},}\ }\href@noop {} {\bibfield
  {journal} {\bibinfo  {journal} {Langmuir}\ }\textbf {\bibinfo {volume}
  {28}},\ \bibinfo {pages} {12426--12431} (\bibinfo {year} {2012})}\BibitemShut
  {NoStop}%
\bibitem [{\citenamefont {Young}, \citenamefont {Goldstein},\ and\
  \citenamefont {Block}(1959)}]{young1959motion}%
  \BibitemOpen
  \bibfield  {author} {\bibinfo {author} {\bibfnamefont {N.}~\bibnamefont
  {Young}}, \bibinfo {author} {\bibfnamefont {J.~S.}\ \bibnamefont
  {Goldstein}}, \ and\ \bibinfo {author} {\bibfnamefont {M.~J.}\ \bibnamefont
  {Block}},\ }\bibfield  {title} {\enquote {\bibinfo {title} {The motion of
  bubbles in a vertical temperature gradient},}\ }\href@noop {} {\bibfield
  {journal} {\bibinfo  {journal} {Journal of Fluid Mechanics}\ }\textbf
  {\bibinfo {volume} {6}},\ \bibinfo {pages} {350--356} (\bibinfo {year}
  {1959})}\BibitemShut {NoStop}%
\bibitem [{\citenamefont {Levich}\ and\ \citenamefont
  {Kuznetsov}(1962)}]{levich1962motion}%
  \BibitemOpen
  \bibfield  {author} {\bibinfo {author} {\bibfnamefont {V.~G.}\ \bibnamefont
  {Levich}}\ and\ \bibinfo {author} {\bibfnamefont {A.~M.}\ \bibnamefont
  {Kuznetsov}},\ }\bibfield  {title} {\enquote {\bibinfo {title} {Motion of
  drops in liquids under the influence of surface-active substances},}\ }in\
  \href@noop {} {\emph {\bibinfo {booktitle} {Doklady Akademii Nauk}}},\ Vol.\
  \bibinfo {volume} {146}\ (\bibinfo {organization} {Russian Academy of
  Sciences},\ \bibinfo {year} {1962})\ pp.\ \bibinfo {pages}
  {145--147}\BibitemShut {NoStop}%
\bibitem [{\citenamefont {Izri}\ \emph {et~al.}(2014)\citenamefont {Izri},
  \citenamefont {Van Der~Linden}, \citenamefont {Michelin},\ and\ \citenamefont
  {Dauchot}}]{izri2014self}%
  \BibitemOpen
  \bibfield  {author} {\bibinfo {author} {\bibfnamefont {Z.}~\bibnamefont
  {Izri}}, \bibinfo {author} {\bibfnamefont {M.~N.}\ \bibnamefont {Van
  Der~Linden}}, \bibinfo {author} {\bibfnamefont {S.}~\bibnamefont {Michelin}},
  \ and\ \bibinfo {author} {\bibfnamefont {O.}~\bibnamefont {Dauchot}},\
  }\bibfield  {title} {\enquote {\bibinfo {title} {Self-propulsion of pure
  water droplets by spontaneous marangoni-stress-driven motion},}\ }\href@noop
  {} {\bibfield  {journal} {\bibinfo  {journal} {Physical review letters}\
  }\textbf {\bibinfo {volume} {113}},\ \bibinfo {pages} {248302} (\bibinfo
  {year} {2014})}\BibitemShut {NoStop}%
\bibitem [{\citenamefont {Herminghaus}\ \emph {et~al.}(2014)\citenamefont
  {Herminghaus}, \citenamefont {Maass}, \citenamefont {Kr{\"u}ger},
  \citenamefont {Thutupalli}, \citenamefont {Goehring},\ and\ \citenamefont
  {Bahr}}]{herminghaus2014interfacial}%
  \BibitemOpen
  \bibfield  {author} {\bibinfo {author} {\bibfnamefont {S.}~\bibnamefont
  {Herminghaus}}, \bibinfo {author} {\bibfnamefont {C.~C.}\ \bibnamefont
  {Maass}}, \bibinfo {author} {\bibfnamefont {C.}~\bibnamefont {Kr{\"u}ger}},
  \bibinfo {author} {\bibfnamefont {S.}~\bibnamefont {Thutupalli}}, \bibinfo
  {author} {\bibfnamefont {L.}~\bibnamefont {Goehring}}, \ and\ \bibinfo
  {author} {\bibfnamefont {C.}~\bibnamefont {Bahr}},\ }\bibfield  {title}
  {\enquote {\bibinfo {title} {Interfacial mechanisms in active emulsions},}\
  }\href@noop {} {\bibfield  {journal} {\bibinfo  {journal} {Soft matter}\
  }\textbf {\bibinfo {volume} {10}},\ \bibinfo {pages} {7008--7022} (\bibinfo
  {year} {2014})}\BibitemShut {NoStop}%
\bibitem [{\citenamefont {Hanna}\ and\ \citenamefont
  {Vlahovska}(2010)}]{Hanna2010}%
  \BibitemOpen
  \bibfield  {author} {\bibinfo {author} {\bibfnamefont {J.~A.}\ \bibnamefont
  {Hanna}}\ and\ \bibinfo {author} {\bibfnamefont {P.~M.}\ \bibnamefont
  {Vlahovska}},\ }\bibfield  {title} {\enquote {\bibinfo {title}
  {Surfactant-induced migration of a spherical drop in stokes flow},}\
  }\href@noop {} {\bibfield  {journal} {\bibinfo  {journal} {Physics of
  Fluids}\ }\textbf {\bibinfo {volume} {22}},\ \bibinfo {pages} {013102}
  (\bibinfo {year} {2010})}\BibitemShut {NoStop}%
\bibitem [{\citenamefont {Pak}, \citenamefont {Feng},\ and\ \citenamefont
  {Stone}(2014)}]{pak2014viscous}%
  \BibitemOpen
  \bibfield  {author} {\bibinfo {author} {\bibfnamefont {O.~S.}\ \bibnamefont
  {Pak}}, \bibinfo {author} {\bibfnamefont {J.}~\bibnamefont {Feng}}, \ and\
  \bibinfo {author} {\bibfnamefont {H.~A.}\ \bibnamefont {Stone}},\ }\bibfield
  {title} {\enquote {\bibinfo {title} {Viscous marangoni migration of a drop in
  a poiseuille flow at low surface p{\'e}clet numbers},}\ }\href@noop {}
  {\bibfield  {journal} {\bibinfo  {journal} {Journal of fluid mechanics}\
  }\textbf {\bibinfo {volume} {753}},\ \bibinfo {pages} {535} (\bibinfo {year}
  {2014})}\BibitemShut {NoStop}%
\bibitem [{\citenamefont {Mandal}, \citenamefont {Bandopadhyay},\ and\
  \citenamefont {Chakraborty}(2015)}]{mandal2015effect}%
  \BibitemOpen
  \bibfield  {author} {\bibinfo {author} {\bibfnamefont {S.}~\bibnamefont
  {Mandal}}, \bibinfo {author} {\bibfnamefont {A.}~\bibnamefont
  {Bandopadhyay}}, \ and\ \bibinfo {author} {\bibfnamefont {S.}~\bibnamefont
  {Chakraborty}},\ }\bibfield  {title} {\enquote {\bibinfo {title} {Effect of
  interfacial slip on the cross-stream migration of a drop in an unbounded
  poiseuille flow},}\ }\href@noop {} {\bibfield  {journal} {\bibinfo  {journal}
  {Physical Review E}\ }\textbf {\bibinfo {volume} {92}},\ \bibinfo {pages}
  {023002} (\bibinfo {year} {2015})}\BibitemShut {NoStop}%
\bibitem [{\citenamefont {Thutupalli}, \citenamefont {Seemann},\ and\
  \citenamefont {Herminghaus}(2011)}]{Thutupalli2011}%
  \BibitemOpen
  \bibfield  {author} {\bibinfo {author} {\bibfnamefont {S.}~\bibnamefont
  {Thutupalli}}, \bibinfo {author} {\bibfnamefont {R.}~\bibnamefont {Seemann}},
  \ and\ \bibinfo {author} {\bibfnamefont {S.}~\bibnamefont {Herminghaus}},\
  }\bibfield  {title} {\enquote {\bibinfo {title} {Swarming behavior of simple
  model squirmers},}\ }\href@noop {} {\bibfield  {journal} {\bibinfo  {journal}
  {New Journal of Physics}\ }\textbf {\bibinfo {volume} {13}},\ \bibinfo
  {pages} {073021} (\bibinfo {year} {2011})}\BibitemShut {NoStop}%
\bibitem [{\citenamefont {Tanabe}, \citenamefont {Ogasawara},\ and\
  \citenamefont {Suematsu}(2020)}]{tanabe2020effect}%
  \BibitemOpen
  \bibfield  {author} {\bibinfo {author} {\bibfnamefont {T.}~\bibnamefont
  {Tanabe}}, \bibinfo {author} {\bibfnamefont {T.}~\bibnamefont {Ogasawara}}, \
  and\ \bibinfo {author} {\bibfnamefont {N.~J.}\ \bibnamefont {Suematsu}},\
  }\bibfield  {title} {\enquote {\bibinfo {title} {Effect of a product on
  spontaneous droplet motion driven by a chemical reaction of surfactant},}\
  }\href@noop {} {\bibfield  {journal} {\bibinfo  {journal} {Physical Review
  E}\ }\textbf {\bibinfo {volume} {102}},\ \bibinfo {pages} {023102} (\bibinfo
  {year} {2020})}\BibitemShut {NoStop}%
\bibitem [{\citenamefont {Pawar}\ and\ \citenamefont
  {Stebe}(1996)}]{Pawar1996}%
  \BibitemOpen
  \bibfield  {author} {\bibinfo {author} {\bibfnamefont {Y.}~\bibnamefont
  {Pawar}}\ and\ \bibinfo {author} {\bibfnamefont {K.~J.}\ \bibnamefont
  {Stebe}},\ }\bibfield  {title} {\enquote {\bibinfo {title} {Marangoni effects
  on drop deformation in an extensional flow: The role of surfactant physical
  chemistry. i. insoluble surfactants},}\ }\href@noop {} {\bibfield  {journal}
  {\bibinfo  {journal} {Physics of Fluids}\ }\textbf {\bibinfo {volume} {8}},\
  \bibinfo {pages} {1738--1751} (\bibinfo {year} {1996})}\BibitemShut {NoStop}%
\bibitem [{\citenamefont {Adamson}, \citenamefont {Adamson},\ and\
  \citenamefont {Gast}(1997)}]{Adamson1997}%
  \BibitemOpen
  \bibfield  {author} {\bibinfo {author} {\bibfnamefont {T.}~\bibnamefont
  {Adamson}}, \bibinfo {author} {\bibfnamefont {A.}~\bibnamefont {Adamson}}, \
  and\ \bibinfo {author} {\bibfnamefont {A.}~\bibnamefont {Gast}},\ }\href
  {https://books.google.co.in/books?id=rK4PAQAAMAAJ} {\emph {\bibinfo {title}
  {Physical Chemistry of Surfaces}}},\ A Wiley-Interscience publication\
  (\bibinfo  {publisher} {Wiley},\ \bibinfo {year} {1997})\BibitemShut
  {NoStop}%
\bibitem [{\citenamefont {Stone}(1990)}]{Stone1990}%
  \BibitemOpen
  \bibfield  {author} {\bibinfo {author} {\bibfnamefont {H.}~\bibnamefont
  {Stone}},\ }\bibfield  {title} {\enquote {\bibinfo {title} {A simple
  derivation of the time-dependent convective-diffusion equation for surfactant
  transport along a deforming interface},}\ }\href@noop {} {\bibfield
  {journal} {\bibinfo  {journal} {Physics of Fluids A: Fluid Dynamics}\
  }\textbf {\bibinfo {volume} {2}},\ \bibinfo {pages} {111--112} (\bibinfo
  {year} {1990})}\BibitemShut {NoStop}%
\bibitem [{\citenamefont {Leal}(1980)}]{leal1980particle}%
  \BibitemOpen
  \bibfield  {author} {\bibinfo {author} {\bibfnamefont {L.}~\bibnamefont
  {Leal}},\ }\bibfield  {title} {\enquote {\bibinfo {title} {Particle motions
  in a viscous fluid},}\ }\href@noop {} {\bibfield  {journal} {\bibinfo
  {journal} {Annual Review of Fluid Mechanics}\ }\textbf {\bibinfo {volume}
  {12}},\ \bibinfo {pages} {435--476} (\bibinfo {year} {1980})}\BibitemShut
  {NoStop}%
\bibitem [{\citenamefont {Nadim}, \citenamefont {Haj-Hariri},\ and\
  \citenamefont {Borhan}(1990)}]{nadim1990thermocapillary}%
  \BibitemOpen
  \bibfield  {author} {\bibinfo {author} {\bibfnamefont {A.}~\bibnamefont
  {Nadim}}, \bibinfo {author} {\bibfnamefont {H.}~\bibnamefont {Haj-Hariri}}, \
  and\ \bibinfo {author} {\bibfnamefont {A.}~\bibnamefont {Borhan}},\
  }\bibfield  {title} {\enquote {\bibinfo {title} {Thermocapillary migration of
  slightly deformed droplets},}\ }\href@noop {} {\bibfield  {journal} {\bibinfo
   {journal} {Particulate science and technology}\ }\textbf {\bibinfo {volume}
  {8}},\ \bibinfo {pages} {191--198} (\bibinfo {year} {1990})}\BibitemShut
  {NoStop}%
\bibitem [{\citenamefont {Rallison}(1978)}]{rallison1978note}%
  \BibitemOpen
  \bibfield  {author} {\bibinfo {author} {\bibfnamefont {J.}~\bibnamefont
  {Rallison}},\ }\bibfield  {title} {\enquote {\bibinfo {title} {Note on the
  fax{\'e}n relations for a particle in stokes flow},}\ }\href@noop {}
  {\bibfield  {journal} {\bibinfo  {journal} {Journal of Fluid Mechanics}\
  }\textbf {\bibinfo {volume} {88}},\ \bibinfo {pages} {529--533} (\bibinfo
  {year} {1978})}\BibitemShut {NoStop}%
\bibitem [{\citenamefont {Hetsroni}\ and\ \citenamefont
  {Haber}(1970)}]{Hetsroni1970}%
  \BibitemOpen
  \bibfield  {author} {\bibinfo {author} {\bibfnamefont {G.}~\bibnamefont
  {Hetsroni}}\ and\ \bibinfo {author} {\bibfnamefont {S.}~\bibnamefont
  {Haber}},\ }\bibfield  {title} {\enquote {\bibinfo {title} {The flow in and
  around a droplet or bubble submerged in an unbound arbitrary velocity
  field},}\ }\href@noop {} {\bibfield  {journal} {\bibinfo  {journal}
  {Rheologica Acta}\ }\textbf {\bibinfo {volume} {9}},\ \bibinfo {pages}
  {488--496} (\bibinfo {year} {1970})}\BibitemShut {NoStop}%
\bibitem [{\citenamefont {Nadim}\ and\ \citenamefont
  {Stone}(1991)}]{Nadim1991}%
  \BibitemOpen
  \bibfield  {author} {\bibinfo {author} {\bibfnamefont {A.}~\bibnamefont
  {Nadim}}\ and\ \bibinfo {author} {\bibfnamefont {H.~A.}\ \bibnamefont
  {Stone}},\ }\bibfield  {title} {\enquote {\bibinfo {title} {The motion of
  small particles and droplets in quadratic flows},}\ }\href@noop {} {\bibfield
   {journal} {\bibinfo  {journal} {Studies in Applied Mathematics}\ }\textbf
  {\bibinfo {volume} {85}},\ \bibinfo {pages} {53--73} (\bibinfo {year}
  {1991})}\BibitemShut {NoStop}%
\bibitem [{\citenamefont {Levich}(1962)}]{levich1962physicochemical}%
  \BibitemOpen
  \bibfield  {author} {\bibinfo {author} {\bibfnamefont {V.~G.}\ \bibnamefont
  {Levich}},\ }\bibfield  {title} {\enquote {\bibinfo {title} {Physicochemical
  hydrodynamics},}\ }\href@noop {} {\  (\bibinfo {year} {1962})}\BibitemShut
  {NoStop}%
\bibitem [{\citenamefont {Palaniappan}\ \emph {et~al.}(1992)\citenamefont
  {Palaniappan}, \citenamefont {Nigam}, \citenamefont {Amaranath},\ and\
  \citenamefont {Usha}}]{Palaniappan1992}%
  \BibitemOpen
  \bibfield  {author} {\bibinfo {author} {\bibfnamefont {D.}~\bibnamefont
  {Palaniappan}}, \bibinfo {author} {\bibfnamefont {S.}~\bibnamefont {Nigam}},
  \bibinfo {author} {\bibfnamefont {T.}~\bibnamefont {Amaranath}}, \ and\
  \bibinfo {author} {\bibfnamefont {R.}~\bibnamefont {Usha}},\ }\bibfield
  {title} {\enquote {\bibinfo {title} {Lamb's solution of stokes's equations: a
  sphere theorem},}\ }\href@noop {} {\bibfield  {journal} {\bibinfo  {journal}
  {The Quarterly Journal of Mechanics and Applied Mathematics}\ }\textbf
  {\bibinfo {volume} {45}},\ \bibinfo {pages} {47--56} (\bibinfo {year}
  {1992})}\BibitemShut {NoStop}%
\bibitem [{\citenamefont {Choudhuri}\ and\ \citenamefont
  {Raja~Sekhar}(2013)}]{Choudhuri2013}%
  \BibitemOpen
  \bibfield  {author} {\bibinfo {author} {\bibfnamefont {D.}~\bibnamefont
  {Choudhuri}}\ and\ \bibinfo {author} {\bibfnamefont {G.}~\bibnamefont
  {Raja~Sekhar}},\ }\bibfield  {title} {\enquote {\bibinfo {title}
  {Thermocapillary drift on a spherical drop in a viscous fluid},}\ }\href@noop
  {} {\bibfield  {journal} {\bibinfo  {journal} {Physics of Fluids}\ }\textbf
  {\bibinfo {volume} {25}},\ \bibinfo {pages} {043104} (\bibinfo {year}
  {2013})}\BibitemShut {NoStop}%
\bibitem [{\citenamefont {Leal}(2007)}]{leal2007advanced}%
  \BibitemOpen
  \bibfield  {author} {\bibinfo {author} {\bibfnamefont {L.~G.}\ \bibnamefont
  {Leal}},\ }\href@noop {} {\emph {\bibinfo {title} {Advanced transport
  phenomena: fluid mechanics and convective transport processes}}},\
  Vol.~\bibinfo {volume} {7}\ (\bibinfo  {publisher} {Cambridge University
  Press},\ \bibinfo {year} {2007})\BibitemShut {NoStop}%
\bibitem [{\citenamefont {Mandal}, \citenamefont {Bandopadhyay},\ and\
  \citenamefont {Chakraborty}(2016)}]{mandal_bandopadhyay_chakraborty_2016}%
  \BibitemOpen
  \bibfield  {author} {\bibinfo {author} {\bibfnamefont {S.}~\bibnamefont
  {Mandal}}, \bibinfo {author} {\bibfnamefont {A.}~\bibnamefont
  {Bandopadhyay}}, \ and\ \bibinfo {author} {\bibfnamefont {S.}~\bibnamefont
  {Chakraborty}},\ }\bibfield  {title} {\enquote {\bibinfo {title} {The effect
  of uniform electric field on the cross-stream migration of a drop in plane
  poiseuille flow},}\ }\href {\doibase 10.1017/jfm.2016.677} {\bibfield
  {journal} {\bibinfo  {journal} {Journal of Fluid Mechanics}\ }\textbf
  {\bibinfo {volume} {809}},\ \bibinfo {pages} {726–774} (\bibinfo {year}
  {2016})}\BibitemShut {NoStop}%
\bibitem [{\citenamefont {Takagi}\ and\ \citenamefont
  {Matsumoto}(2011)}]{takagi2011surfactant}%
  \BibitemOpen
  \bibfield  {author} {\bibinfo {author} {\bibfnamefont {S.}~\bibnamefont
  {Takagi}}\ and\ \bibinfo {author} {\bibfnamefont {Y.}~\bibnamefont
  {Matsumoto}},\ }\bibfield  {title} {\enquote {\bibinfo {title} {Surfactant
  effects on bubble motion and bubbly flows},}\ }\href@noop {} {\bibfield
  {journal} {\bibinfo  {journal} {Annual Review of Fluid Mechanics}\ }\textbf
  {\bibinfo {volume} {43}},\ \bibinfo {pages} {615--636} (\bibinfo {year}
  {2011})}\BibitemShut {NoStop}%
\bibitem [{\citenamefont {Di~Carlo}\ \emph {et~al.}(2007)\citenamefont
  {Di~Carlo}, \citenamefont {Irimia}, \citenamefont {Tompkins},\ and\
  \citenamefont {Toner}}]{di2007continuous}%
  \BibitemOpen
  \bibfield  {author} {\bibinfo {author} {\bibfnamefont {D.}~\bibnamefont
  {Di~Carlo}}, \bibinfo {author} {\bibfnamefont {D.}~\bibnamefont {Irimia}},
  \bibinfo {author} {\bibfnamefont {R.~G.}\ \bibnamefont {Tompkins}}, \ and\
  \bibinfo {author} {\bibfnamefont {M.}~\bibnamefont {Toner}},\ }\bibfield
  {title} {\enquote {\bibinfo {title} {Continuous inertial focusing, ordering,
  and separation of particles in microchannels},}\ }\href@noop {} {\bibfield
  {journal} {\bibinfo  {journal} {Proceedings of the National Academy of
  Sciences}\ }\textbf {\bibinfo {volume} {104}},\ \bibinfo {pages}
  {18892--18897} (\bibinfo {year} {2007})}\BibitemShut {NoStop}%
\bibitem [{\citenamefont {Seemann}\ \emph {et~al.}(2011)\citenamefont
  {Seemann}, \citenamefont {Brinkmann}, \citenamefont {Pfohl},\ and\
  \citenamefont {Herminghaus}}]{seemann2011droplet}%
  \BibitemOpen
  \bibfield  {author} {\bibinfo {author} {\bibfnamefont {R.}~\bibnamefont
  {Seemann}}, \bibinfo {author} {\bibfnamefont {M.}~\bibnamefont {Brinkmann}},
  \bibinfo {author} {\bibfnamefont {T.}~\bibnamefont {Pfohl}}, \ and\ \bibinfo
  {author} {\bibfnamefont {S.}~\bibnamefont {Herminghaus}},\ }\bibfield
  {title} {\enquote {\bibinfo {title} {Droplet based microfluidics},}\
  }\href@noop {} {\bibfield  {journal} {\bibinfo  {journal} {Reports on
  progress in physics}\ }\textbf {\bibinfo {volume} {75}},\ \bibinfo {pages}
  {016601} (\bibinfo {year} {2011})}\BibitemShut {NoStop}%
\bibitem [{\citenamefont {Schramm}\ \emph {et~al.}(1992)\citenamefont {Schramm}
  \emph {et~al.}}]{schramm1992fundamentals}%
  \BibitemOpen
  \bibfield  {author} {\bibinfo {author} {\bibfnamefont {L.~L.}\ \bibnamefont
  {Schramm}} \emph {et~al.},\ }\bibfield  {title} {\enquote {\bibinfo {title}
  {Fundamentals and applications in the petroleum industry},}\ }\href@noop {}
  {\bibfield  {journal} {\bibinfo  {journal} {Adv. Chem}\ }\textbf {\bibinfo
  {volume} {231}},\ \bibinfo {pages} {3--24} (\bibinfo {year}
  {1992})}\BibitemShut {NoStop}%
\bibitem [{\citenamefont {Teh}\ \emph {et~al.}(2008)\citenamefont {Teh},
  \citenamefont {Lin}, \citenamefont {Hung},\ and\ \citenamefont
  {Lee}}]{teh2008droplet}%
  \BibitemOpen
  \bibfield  {author} {\bibinfo {author} {\bibfnamefont {S.-Y.}\ \bibnamefont
  {Teh}}, \bibinfo {author} {\bibfnamefont {R.}~\bibnamefont {Lin}}, \bibinfo
  {author} {\bibfnamefont {L.-H.}\ \bibnamefont {Hung}}, \ and\ \bibinfo
  {author} {\bibfnamefont {A.~P.}\ \bibnamefont {Lee}},\ }\bibfield  {title}
  {\enquote {\bibinfo {title} {Droplet microfluidics},}\ }\href@noop {}
  {\bibfield  {journal} {\bibinfo  {journal} {Lab on a Chip}\ }\textbf
  {\bibinfo {volume} {8}},\ \bibinfo {pages} {198--220} (\bibinfo {year}
  {2008})}\BibitemShut {NoStop}%
\bibitem [{\citenamefont {Chan}\ and\ \citenamefont
  {Leal}(1979)}]{chan1979motion}%
  \BibitemOpen
  \bibfield  {author} {\bibinfo {author} {\bibfnamefont {P.-H.}\ \bibnamefont
  {Chan}}\ and\ \bibinfo {author} {\bibfnamefont {L.}~\bibnamefont {Leal}},\
  }\bibfield  {title} {\enquote {\bibinfo {title} {The motion of a deformable
  drop in a second-order fluid},}\ }\href@noop {} {\bibfield  {journal}
  {\bibinfo  {journal} {Journal of Fluid Mechanics}\ }\textbf {\bibinfo
  {volume} {92}},\ \bibinfo {pages} {131--170} (\bibinfo {year}
  {1979})}\BibitemShut {NoStop}%
\bibitem [{\citenamefont {Li}\ and\ \citenamefont
  {Sarkar}(2005)}]{li2005effects}%
  \BibitemOpen
  \bibfield  {author} {\bibinfo {author} {\bibfnamefont {X.}~\bibnamefont
  {Li}}\ and\ \bibinfo {author} {\bibfnamefont {K.}~\bibnamefont {Sarkar}},\
  }\bibfield  {title} {\enquote {\bibinfo {title} {Effects of inertia on the
  rheology of a dilute emulsion of drops in shear},}\ }\href@noop {} {\bibfield
   {journal} {\bibinfo  {journal} {Journal of Rheology}\ }\textbf {\bibinfo
  {volume} {49}},\ \bibinfo {pages} {1377--1394} (\bibinfo {year}
  {2005})}\BibitemShut {NoStop}%
\bibitem [{\citenamefont {Mukherjee}\ and\ \citenamefont
  {Sarkar}(2014)}]{mukherjee2014lateral}%
  \BibitemOpen
  \bibfield  {author} {\bibinfo {author} {\bibfnamefont {S.}~\bibnamefont
  {Mukherjee}}\ and\ \bibinfo {author} {\bibfnamefont {K.}~\bibnamefont
  {Sarkar}},\ }\bibfield  {title} {\enquote {\bibinfo {title} {Lateral
  migration of a viscoelastic drop in a newtonian fluid in a shear flow near a
  wall},}\ }\href@noop {} {\bibfield  {journal} {\bibinfo  {journal} {Physics
  of fluids}\ }\textbf {\bibinfo {volume} {26}},\ \bibinfo {pages} {103102}
  (\bibinfo {year} {2014})}\BibitemShut {NoStop}%
\bibitem [{\citenamefont {Bhagat}\ \emph {et~al.}(2010)\citenamefont {Bhagat},
  \citenamefont {Bow}, \citenamefont {Hou}, \citenamefont {Tan}, \citenamefont
  {Han},\ and\ \citenamefont {Lim}}]{bhagat2010microfluidics}%
  \BibitemOpen
  \bibfield  {author} {\bibinfo {author} {\bibfnamefont {A.~A.~S.}\
  \bibnamefont {Bhagat}}, \bibinfo {author} {\bibfnamefont {H.}~\bibnamefont
  {Bow}}, \bibinfo {author} {\bibfnamefont {H.~W.}\ \bibnamefont {Hou}},
  \bibinfo {author} {\bibfnamefont {S.~J.}\ \bibnamefont {Tan}}, \bibinfo
  {author} {\bibfnamefont {J.}~\bibnamefont {Han}}, \ and\ \bibinfo {author}
  {\bibfnamefont {C.~T.}\ \bibnamefont {Lim}},\ }\bibfield  {title} {\enquote
  {\bibinfo {title} {Microfluidics for cell separation},}\ }\href@noop {}
  {\bibfield  {journal} {\bibinfo  {journal} {Medical \& biological engineering
  \& computing}\ }\textbf {\bibinfo {volume} {48}},\ \bibinfo {pages}
  {999--1014} (\bibinfo {year} {2010})}\BibitemShut {NoStop}%
\bibitem [{\citenamefont {Zhu}\ \emph {et~al.}(2014)\citenamefont {Zhu},
  \citenamefont {Zhu}, \citenamefont {Guo}, \citenamefont {Cui}, \citenamefont
  {Ye},\ and\ \citenamefont {Fang}}]{zhu2014nanoliter}%
  \BibitemOpen
  \bibfield  {author} {\bibinfo {author} {\bibfnamefont {Y.}~\bibnamefont
  {Zhu}}, \bibinfo {author} {\bibfnamefont {L.-N.}\ \bibnamefont {Zhu}},
  \bibinfo {author} {\bibfnamefont {R.}~\bibnamefont {Guo}}, \bibinfo {author}
  {\bibfnamefont {H.-J.}\ \bibnamefont {Cui}}, \bibinfo {author} {\bibfnamefont
  {S.}~\bibnamefont {Ye}}, \ and\ \bibinfo {author} {\bibfnamefont
  {Q.}~\bibnamefont {Fang}},\ }\bibfield  {title} {\enquote {\bibinfo {title}
  {Nanoliter-scale protein crystallization and screening with a microfluidic
  droplet robot},}\ }\href@noop {} {\bibfield  {journal} {\bibinfo  {journal}
  {Scientific reports}\ }\textbf {\bibinfo {volume} {4}},\ \bibinfo {pages}
  {1--9} (\bibinfo {year} {2014})}\BibitemShut {NoStop}%
\bibitem [{\citenamefont {Pillai}\ and\ \citenamefont
  {Narayanan}(2018)}]{pillai2018rayleigh}%
  \BibitemOpen
  \bibfield  {author} {\bibinfo {author} {\bibfnamefont {D.~S.}\ \bibnamefont
  {Pillai}}\ and\ \bibinfo {author} {\bibfnamefont {R.}~\bibnamefont
  {Narayanan}},\ }\bibfield  {title} {\enquote {\bibinfo {title}
  {Rayleigh--taylor stability in an evaporating binary mixture},}\ }\href@noop
  {} {\bibfield  {journal} {\bibinfo  {journal} {Journal of Fluid Mechanics}\
  }\textbf {\bibinfo {volume} {848}} (\bibinfo {year} {2018})}\BibitemShut
  {NoStop}%
\end{thebibliography}%

\end{document}